\documentclass[aps,prd,10pt,showpacs,amsmath,showkeys,twocolumn,floatfix,amssymb, preprintnumbers, nofootinbib, superscriptaddress]{revtex4-1} 
\usepackage{epsfig,dcolumn}
\usepackage{graphicx}
\usepackage{comment} 
\usepackage[usenames]{color}
\usepackage{graphicx}
\usepackage{bm}
 \usepackage{ifpdf}
  \usepackage{floatrow}
\usepackage{makecell}
 \usepackage{caption}

\usepackage[normalem]{ulem}
\usepackage[dvipsnames]{xcolor}
\usepackage[utf8]{inputenc}
\usepackage{hyperref}
\hypersetup{ 
  pdfnewwindow=true,      % links in new window
  colorlinks=true,        % false: boxed links; true: colored links
  linkcolor=PineGreen,    % color of internal links
  citecolor=PineGreen,    % color of links to bibliography
  filecolor=PineGreen,    % color of file links
  urlcolor=PineGreen      % color of external links
}

\newcommand{\be}{\begin{eqnarray}}
\newcommand{\ee}{\end{eqnarray}}

 \usepackage{epsfig,dcolumn}
\usepackage{graphicx}
\usepackage{comment} 
\usepackage[usenames]{color}
\usepackage{bm}
\usepackage{ifpdf}
\usepackage{floatrow}
\usepackage{makecell}
\usepackage{caption}
 
\begin{document}

\title{Modeling few-body  resonances   in finite volume}

\author{Peng~Guo}
\email{pguo@csub.edu}

\affiliation{Department of Physics and Engineering,  California State University, Bakersfield, CA 93311, USA}
\affiliation{Kavli Institute for Theoretical Physics, University of California, Santa Barbara, CA 93106, USA}

\date{\today}

\begin{abstract} 
Under the assumption of separable interactions, we illustrate how the few-body quantization condition may be formulated in terms of phase shifts in general, which may be useful for describing and modeling of few-body resonances   in finite volume.
\end{abstract}

\maketitle

\section{Introduction}\label{intro}

Few-hadron dynamics plays an important role in hadron and nuclear physics. There have been many good examples of physics processes that can only be  understood  through few-body interactions, such as,   the   $u$- and $d$-quark mass difference in  \mbox{$\eta \rightarrow 3 \pi$}   \cite{Kambor:1995yc,Anisovich:1996tx,Schneider:2010hs,Kampf:2011wr,Guo:2015zqa,Guo:2016wsi,Colangelo:2016jmc},     Efimov states  \cite{Efimov:1970zz, Braaten:2004rn} and halo nuclei  \cite{Zhukov:1993aw,Hammer:2017tjm}. The understanding of few-body interaction is also crucial in recent experimental efforts of exotic hadrons study, since most of exotic hadron states are expected to appear as few-hadron resonances.  On the theory side,   lattice Quantum Chromodynamics (LQCD) provides an   ab-initio  method for the study of exotic hadron states. However,    LQCD computation is usually  performed   in Euclidean space with  certain  periodic boundary condition, normally  only   discrete energy spectrum are measured  in numerical simulation.  Hence, mapping out few-hadron dynamics from discrete energy spectrum is a key step for the study of exotic hadron states in LQCD.
In   two-body sector,  L\"uscher  formula \cite{Luscher:1990ux} and its  variants  \cite{Rummukainen:1995vs,Christ:2005gi,Bernard:2008ax,He:2005ey,Lage:2009zv,Doring:2011vk,Briceno:2012yi,Hansen:2012tf,Guo:2012hv,Guo:2013vsa} provide an elegant form of    mapping   out two-body phase shift  from discrete energy levels.

In past few years, many   progresses  from different approaches \cite{Kreuzer:2008bi,Kreuzer:2009jp,Kreuzer:2012sr,Polejaeva:2012ut,Briceno:2012rv,Hansen:2014eka,Hansen:2015zga,Hansen:2016fzj,Briceno:2017tce,Hammer:2017uqm,Hammer:2017kms,Meissner:2014dea,Mai:2017bge,Mai:2018djl, Doring:2018xxx, Romero-Lopez:2018rcb,Guo:2016fgl,Guo:2017ism,Guo:2017crd,Guo:2018xbv,Blanton:2019igq,Romero-Lopez:2019qrt,Blanton:2019vdk,Mai:2019fba,Guo:2018ibd,Guo:2019hih,Guo:2019ogp,Guo:2020wbl,Guo:2020kph,Guo:2020iep,Guo:2020ikh,Hansen:2020zhy}  have been  made going beyond three-body threshold. Although few-body quantization conditions   are formulated   differently among these groups, it has been very clear \cite{Guo:2020iep} that in few-body sectors, the few-body   amplitudes  are not directly extracted from lattice results. Particle interactions or its associated subprocess   amplitudes are in fact essential ingredients in quantization condition. The infinite volume few-body amplitudes that are generated by particle interactions through coupled integral equations    must be computed in a separate step once these dynamical ingredients are determined.  In order to make predictions or fit lattice results,   dynamical ingredients of quantization condition, such as interaction potentials or off-shell subprocess amplitudes must be modeled one way or another. In addition,    number of partial waves  involved in some physical processes may be large,  which may add some extra complications on top of the uncertainty in modeling itself.  Therefore, to have a reliable and controllable predictions, the modeling of dynamical ingredients must be constrained or guided by experimental data or effective theory. Nevertheless,  there are two physical regions in which    predictions  and calculations may be made fairly reliable:    (1) near   threshold      which is the region where  the physical  reaction can be described rather precisely by  non-relativistic potential theory or relativistic effective perturbation theory;     (2) near resonances region where  resonance properties may be less affected by modeling and other partial waves.

In present work, we  focus on near resonance region and aim to provide an approximate means for the modeling of few-body resonances in finite volume.   Based on separable interaction potential assumption,  we illustrate how  the few-body  quantization condition may be formulated in terms of subprocess phase shifts. Hence, the resonances may be modeled and inserted into quantization condition  through phase shifts.   Both two-body and three-body subprocess amplitudes   appear     L\"uscher  formula-like and   solutions are given by algebra equations. The aim of this work is to illustrate a simple  way of parameterization of dynamics of few-body resonance in finite volume. Under the separable short-range potential approximation,  the few-body formalism is much simplified with  the trade-off  that the approximation may be only valid for the description of sharp resonances dynamics. As a simple illustration,   the formalism is only presented in non-relativistic kinematics, the extension to relativistic kinematics may be possible by  replacing non-relativistic few-body propagators with relativistic ones, see  \cite{Guo:2019ogp,Guo:2020kph}. The relativistic extension of formalism will be presented in future publications.

The paper is organized as follows.   With separable interactions approximation, the technical details of  formulating quantization conditions in terms of phase shifts   are presented    in Sec.~\ref{quantization}.      A summary  is  given in Sec.~\ref{summ}.

\section{Quantization condition under separable interactions assumption}\label{quantization}

Few-body   quantization condition in finite volume can be formulated from homogeneous Faddeev type equations, see \cite{Guo:2020wbl,Guo:2020kph,Guo:2020iep,Guo:2020ikh}. As a simple example, we consider three non-relativistic identical bosons of mass $m$ interacting with both pair-wise interaction and three-body force  in follows. Due to exchange symmetry, only two independent Faddeev amplitudes are required:      $\mathcal{T}^{(2b)}  $ and  $\mathcal{T}^{(3b)}  $ that are associated with pair-wise two-body interaction,  $V^{(2b)}$,  and three-body interaction, $V^{(3b)}$,  by
 \begin{equation}
  \mathcal{T}^{(2b,3b)}(\mathbf{ k}_1 , \mathbf{ k}_2)    =  -  \langle  \mathbf{ k}_1  \mathbf{ k}_2|  m V^{(2b,3b)} | \Psi \rangle ,
 \end{equation}
 where $\Psi$ stands for the three-body total wave function.    The    $(\mathbf{ k}_1 ,\mathbf{ k}_2) \in \frac{2\pi \mathbf{ n}}{L}$,  $\mathbf{ n} \in \mathbb{Z}^3$ refer to particle-1 and -2 momenta respectively, and  third particle momentum is constrained by total momentum conservation, $$\mathbf{ k}_3= - \mathbf{ k}_1  - \mathbf{ k}_2. $$
  In follows, we also use  symbols $(\mathbf{ k}_{13} , \mathbf{ k}_{(13)2} )$  to describe two independent  relative momenta of three particles, where
   \begin{align}
 \mathbf{ k}_{13}  &= \frac{\mathbf{ k}_{1} -\mathbf{ k}_{3}}{2}  = \mathbf{ k}_{1} + \frac{\mathbf{ k}_{2}}{2},  \nonumber \\
 \mathbf{ k}_{(13)2} &  = \sqrt{\frac{1}{3}}  \left ( \frac{\mathbf{ k}_{1} + \mathbf{ k}_{3}   }{2} - \mathbf{ k}_{2} \right ) = - \sqrt{\frac{4}{3}} \mathbf{ k}_{2} .
 \end{align}

The stationary states of  three-body dynamics in finite volume  is  described by homogeneous Faddeev type equations,   see \cite{Guo:2020wbl,Guo:2020kph,Guo:2020iep,Guo:2020ikh},
  \begin{align}
     &  \mathcal{T}^{(2b)} (\mathbf{ k}_1,  \mathbf{ k}_2)    =  -    \frac{1}{L^3}    \sum_{ 
              \mathbf{ p}_{1}    }   \frac{     \tau^{(2b)} (\mathbf{ k}_{13};  \mathbf{ p}_{1} + \frac{\mathbf{ k}_2}{2})     }{mE - \frac{   \mathbf{ p}_{1}^2 + \mathbf{ k}_2^2 +  (\mathbf{ p}_1+ \mathbf{ k}_2)^2   }{2} }    \nonumber \\
              & \quad \quad  \quad \quad  \quad \quad   \times    \bigg [   2   \mathcal{T}^{(2b)} ( \mathbf{ k}_2 , \mathbf{ p}_1)       +   \mathcal{T}^{(3b)} (\mathbf{ p}_1  , \mathbf{ k}_2) \bigg ]  ,  \label{T2bFaddeev}
  \end{align}
 and
   \begin{align}
    &  \mathcal{T}^{(3b)} (\mathbf{ k}_1 , \mathbf{ k}_2)  \nonumber \\
    &   =   -  \frac{1}{L^6}   \sum_{ 
              \mathbf{ p}_{1}   , \mathbf{ p}_{2}    }       \frac{\tau^{(3b)} (\mathbf{ K}  ; \mathbf{ P} ) }{mE - \frac{   \mathbf{ p}_{1}^2 + \mathbf{ p}_2^2 +  (\mathbf{ p}_1+ \mathbf{ p}_2)^2   }{2} }       3   \mathcal{T}^{(2b)} (\mathbf{ p}_1 , \mathbf{ p}_2)    , \label{T3bFaddeev}
 \end{align}
 where  $$ (\mathbf{ p}_{1},\mathbf{ p}_{2} )  \in \frac{2\pi \mathbf{ n}}{L}, \ \ \mathbf{ n} \in \mathbb{Z}^3.$$ 
The symbol $(\mathbf{ K},  \mathbf{ P})$    stand for $6$-dimensional  vectors, they are related to relative momenta $(\mathbf{ k}_{13} , \mathbf{ k}_{(13)2} )$ by
\begin{align}
 \mathbf{ K}  &= \{ \mathbf{ k}_{13} , \mathbf{ k}_{(13)2}  \} = \{ \mathbf{ k}_1 + \frac{\mathbf{ k}_2}{2}  ,- \sqrt{\frac{4}{3}} \mathbf{ k}_{2} \}  , \nonumber \\
 \mathbf{ P} & = \{ \mathbf{ p}_{13} , \mathbf{ p}_{(13)2}  \} = \{ \mathbf{ p}_1 + \frac{\mathbf{ p}_2}{2}  ,- \sqrt{\frac{4}{3}} \mathbf{ p}_{2} \}.
 \end{align}
  The length of $6D$ vectors are given by  
 \begin{align}
 K& =\sqrt{ \mathbf{ k}_{13}^2 + \mathbf{ k}_{(13)2}^2} = \sqrt{ \frac{1}{2} \sum_{i=1}^3 \mathbf{ k}_{i}^2 }, \nonumber \\
  P& =\sqrt{ \mathbf{ p}_{13}^2 + \mathbf{ p}_{(13)2}^2} = \sqrt{ \frac{1}{2} \sum_{i=1}^3 \mathbf{ p}_{i}^2 } .
 \end{align}
Symbols $\tau^{(2b)}$ and  $\tau^{(3b)}$  that are associated with  two-body interaction   $V^{(2b)}$  and three-body interaction  $V^{(3b)}$ respectively  are used to describe   off-shell  subprocess transition   amplitudes between initial and final momenta states.  For example,  $\tau^{(2b)}$  in $(13)$ isobar channel with particle-2 carrying a momentum $\mathbf{ k}_2$  satisfies two-body inhomogeneous Lippmann-Schwinger equations,
\begin{align}
 &    \tau^{(2b)} (\mathbf{ k}_{13};  \mathbf{ k}'_{13})   = - m  \widetilde{V}^{(2b)}   ( | \mathbf{ k}_{13}- \mathbf{ k}'_{13}   |)  \nonumber \\
    &  +   \frac{1}{L^3}    \sum_{ 
              \mathbf{ p}_{1}    }     \frac{m  \widetilde{V}^{(2b)}    ( | \mathbf{ k}_{13} -   \mathbf{ p}_{1} - \frac{\mathbf{ k}_2}{2}    |)  }{mE - \frac{   \mathbf{ p}_{1}^2 + \mathbf{ k}_2^2 + (\mathbf{ p}_1+ \mathbf{ k}_2)^2   }{2} }    \tau^{(2b)} (  \mathbf{ p}_{1} + \frac{\mathbf{ k}_2}{2} ;  \mathbf{ k}'_{13})     ,   \label{tau2beq}
 \end{align}
  and similarly    $\tau^{(3b)}$ satisfies a three-body equation,
   \begin{align}
&   \tau^{(3b)} (\mathbf{ K}  ;   \mathbf{ K}' ) = -m  \widetilde{V}^{(3b)} (| \mathbf{ K} -\mathbf{ K}' | )   \nonumber \\
  &  +   \frac{1}{L^6}    \sum_{ 
              \mathbf{ p}_{1}  ,  \mathbf{ p}_{2}    }  \frac{ m  \widetilde{V}^{(3b)} ( | \mathbf{ K} - \mathbf{ P}  |)   }{mE - \frac{   \mathbf{ p}_{1}^2 + \mathbf{ p}_2^2 + (\mathbf{ p}_1+ \mathbf{ p}_2)^2   }{2} }       \tau^{(3b)} (\mathbf{ P} ; \mathbf{ K}'  )      . \label{tau3beq}
 \end{align}
  $\tau^{(2b)}$ and  $\tau^{(3b)}$ are    dynamical input of   finite volume Faddeev equations in Eq.(\ref{T2bFaddeev}) and Eq.(\ref{T3bFaddeev}), and must be solved first.

The quantization condition without cubic irreducible representation projection is   given by
 \begin{align}
       & 0=  \det \bigg[  L^6 \delta_{ \mathbf{ k}_1, \mathbf{ p}_1 } \delta_{ \mathbf{ k}_2, \mathbf{ p}_2 }  +       \frac{ L^3 \delta_{ \mathbf{ k}_2, \mathbf{ p}_1 }      2 \tau^{(2b)} (\mathbf{ k}_{13} ;  \mathbf{ p}_2 + \frac{\mathbf{ k}_2}{2})     }{mE -\frac{3   \mathbf{ k}_2^2 }{4}  -   (\mathbf{ p}_2+\frac{   \mathbf{ k}_2 }{2}   )^2  }     \nonumber \\
       &              -        \frac{3}{L^3}    \sum_{ 
              \mathbf{ p}     }   \frac{     \tau^{(2b)} (\mathbf{ k}_{13} ;  \mathbf{ p} + \frac{\mathbf{ k}_2}{2})   \tau^{(3b)} ( \{ \mathbf{ p} + \frac{\mathbf{ k}_2}{2} , - \sqrt{\frac{4}{3}} \mathbf{ k}_2 \} ;  \mathbf{ P} )  }{ \left [ mE -\frac{3   \mathbf{ k}_2^2 }{4}  -   (\mathbf{ p}+\frac{   \mathbf{ k}_2 }{2}   )^2 \right ] \left (mE - \mathbf{ P}^2 \right ) }    ,
 \end{align}
 where  $\tau^{(2b)}$ and  $\tau^{(3b)}$  in principle are given by the solutions of Eq.(\ref{tau2beq}) and  Eq.(\ref{tau3beq})  respectively. In Sec.\ref{tau2b3bsol}, we will show that   with separable interaction approximation  Eq.(\ref{tau2beq}) and  Eq.(\ref{tau3beq}) may be converted into algebra equations.  Hence, the solutions of $\tau^{(2b)}$ and  $\tau^{(3b)}$ are L\"uscher  formula-like, and can be formulated in terms of conventional two-body phase shifts in $3D$ and  unconventional but mathematically convenient three-body phase shifts in $6D$. We remark that the momentum sum in quantization condition must be regulated in numerical evaluation of discrete energy levels, which are either associate to ultraviolet  divergence or normalization of determinant condition. In current work, since the technical regularization is not our focus of presentation, the specific procedure of regularization has been left-out, we refer interesting readers to Refs. ~\cite{Hansen:2014eka,Hansen:2015zga}.

\subsection{Separable interactions and algebra solutions of $\tau^{(2b)}$ and  $\tau^{(3b)}$}\label{tau2b3bsol} 
Under the assumption of separable short-range potentials for both $V^{(2b)}$  and  $V^{(3b)}$, the partial wave expansion of potentials thus have the forms of
\begin{align}
&  \widetilde{V}^{(2b)}    ( | \mathbf{ k}_{13} -   \mathbf{ k}'_{13}    |)   \nonumber \\
&=  \sum_{L M} Y_{L M} ( \mathbf{\hat{ k}}_{13} ) g^{(2b)}_L (k_{13}) V_L^{(2b)}  g^{(2b)}_L(k'_{13}) Y^*_{L M} ( \mathbf{\hat{ k}}'_{13} ),
\end{align}
and
\begin{align}
&  \widetilde{V}^{(3b)}    ( | \mathbf{ K}  -   \mathbf{ K}      |)   \nonumber \\
&=  \sum_{[J]} Y_{ [J]} ( \mathbf{\hat{ K}} ) g^{(3b)}_{[J]} ( K) V_J^{(3b)} g^{(3b)}_{[J]}(K' ) Y^*_{[J]} ( \mathbf{\hat{ K}}' ),
\end{align}
where  $Y_{L M}( \mathbf{\hat{ k}}_{13} ) $ is $3D$ spherical harmonic function with quantum numbers  $|LM \rangle $ representing orbital angular momentum configurations between particle-1 and -3, while particle-2  acts as a spectator and is not involved in interaction.    $Y_{ [J]} ( \mathbf{\hat{ K}} )$  stands for the $6D$ hyperspherical harmonic basis function, see Refs.~\cite{Ripelle:1983hh,Ripelle:1993gr} and also Appendix \ref{HHexpand},  the quantum numbers $[J]$ represent  a specific angular momentum configuration of three particles with  a total angular momentum-$J$.   $Y_{ [J]} ( \mathbf{\hat{ K}} )$ may be constructed through   two $3D$ spherical harmonic functions.    For  example, considering  a configuration with angular momentum state $| L_{13} M_{13} \rangle$ between particle-1 and -3 coupled with particle-2 in relative angular momentum state $| L_{(13)2} M_{(13)2} \rangle$ into total angular momentum state $|[J]  \rangle = | JM L_{13} L_{(13)2} \rangle$, thus $Y_{ [J]} ( \mathbf{\hat{ K}} )$ is given by 
\begin{align}
& Y_{ [J]} ( \mathbf{\hat{ K}} ) =  \sum_{M_{13}, M_{(13)2}} \langle L_{13} M_{13} , L_{(13)2} M_{(13)2} | JM \rangle \nonumber \\
& \times Y_{L_{13} M_{13}} ( \mathbf{\hat{ k}}_{13} )  Y_{L_{(13)2} M_{(13)2} } ( \mathbf{\hat{ k}}_{(13)2} ) \mathcal{P}_{J L_{13} L_{(13)2}} ( \phi),
\end{align}
where  $$ \phi = \tan^{-1} \frac{k_{13}}{k_{(13)2}}.$$  The   function $ \mathcal{P}_{J L_{13} L_{(13)2}} (\phi)$ is related to Jacobi polynomial by, also see \cite{Ripelle:1983hh,Ripelle:1993gr},
\begin{align}
 \mathcal{P}_{J L_{13} L_{(13)2}} (\phi)  &= N_{J L_{13} L_{(13)2}} (\sin \phi)^{L_{13}} (\cos \phi)^{L_{(13)2}} \nonumber \\
& \times  P^{(L_{13} + \frac{1}{2}, L_{(13)2} + \frac{1}{2})}_{\frac{J- L_{13} -L_{(13)2}}{2}} ( \cos 2 \phi),
\end{align}
the normalization factor $N_{J L_{13} L_{(13)2}}$ is determined by orthonormal relation,
\begin{align}
 \int_0^{\frac{\pi}{2}} d \phi \sin^2 \phi \cos^2 \phi \mathcal{P}_{J L_{13} L_{(13)2}} (\phi)   \mathcal{P}_{J' L_{13} L_{(13)2}} (\phi)  = \delta_{J, J'}.
\end{align}
 The form factors, $g^{(2b)}_L$ and $g^{(3b)}_{[J]} $, and  potential strengths, $V_L^{(2b)}$ and $V_J^{(3b)}$,  may be considered as model parameters. Usually, the form factors, such as  $g^{(2b)}_L$, must show the correct threshold behavior, 
\begin{equation}
g^{(2b)}_L(k \rightarrow 0)  \sim k^L . 
\end{equation}
The potential strengths $V_L^{(2b)}$ and $V_J^{(3b)}$ may be used to model two-body and three-body resonances, for example, the two-particle resonance of mass $m^{(2b)}_R$ in $(13)$  isobar pair  channel with particle-2 carrying momentum $\mathbf{ k}_2$  may be given by
\begin{equation}
V_L^{(2b)}  \propto  \frac{ 1 }{    ( E - \frac{3}{4} \frac{ \mathbf{ k}^2_2}{m } )  -  m^{(2b)}_R}. 
\end{equation}
A three-particle resonance of mass $m^{(3b)}_R$ thus may be modeled  similarly by
\begin{equation}
V_L^{(3b)}  \propto  \frac{ 1 }{    E  -  m^{(3b)}_R}. 
\end{equation}

Separable interactions suggest that $\tau^{(2b)}$ in Eq.(\ref{tau2beq}) and  $\tau^{(3b)}$ in Eq.(\ref{tau3beq})  may be given  by   L\"uscher  formula-like  algebra equations,   see detailed discussion in Appendix \ref{HHexpand},
\begin{align}
&  \frac{\sqrt{mE - \frac{3}{4} \mathbf{ k}^2_2}  }{ 16 \pi^{ 2}    }   \tau^{(2b)} (\mathbf{ k}_{13};  \mathbf{ k}'_{13})   \nonumber \\
&  =  \sum_{L M, L'M'}      \frac{Y_{L M} ( \mathbf{\hat{ k}}_{13} )   g^{(2b)}_L (k_{13})   g^{(2b)}_{L'}(k'_{13}) Y^*_{L' M'} ( \mathbf{\hat{ k}}'_{13} )  }{ g^{(2b)}_L ( \sqrt{mE - \frac{3}{4} \mathbf{ k}^2_2})g^{(2b)}_{L'} ( \sqrt{mE - \frac{3}{4} \mathbf{ k}^2_2}) }   \nonumber \\
& \times    i^{L-L'}   \bigg [   \delta_{L M, L' M'}      \cot \delta^{(2b)}_{L} (\sqrt{mE - \frac{3}{4} \mathbf{ k}^2_2} )  \nonumber \\
& \quad \quad \quad \quad  \quad  \quad       -  \mathcal{M}^{(2b, \mathbf{ k}_2)}_{L M, L' M'} (   \sqrt{mE - \frac{3}{4} \mathbf{ k}^2_2})  \bigg ]^{-1} , \label{tau2bsol}
\end{align}
and
\begin{align}
 &\frac{ (mE)^{2} }{128 \pi^{ 5}  }  \tau^{(3b)} (\mathbf{ K} ;   \mathbf{ K}'  )   \nonumber \\
 & =  \sum_{[J], [J']}  \frac{ Y_{ [J]} ( \mathbf{\hat{ K}} )  g^{(3b)}_{[J]} ( K)  g^{(3b)}_{[J']}(K' )  Y^*_{[J']} ( \mathbf{\hat{ K}}' )}{  g^{(3b)}_{[J]} ( \sqrt{mE} ) g^{(3b)}_{[J']} ( \sqrt{mE} ) }   \nonumber \\
&     \times  i^{J-J'}   \left [  \delta_{[J], [J']}     \cot \delta^{(3b)}_J ( \sqrt{m E})  - \mathcal{M}^{(3b)}_{[J], [J']} (\sqrt{m E})   \right ]^{-1}  . \label{tau3bsol}
\end{align}
 Generalized L\"uscher  zeta functions, $ \mathcal{M}^{(2b, \mathbf{ k}_2)} $ in $3D$ and  $\mathcal{M}^{(3b)}$  in $6D$,      are given respectively  by 
\begin{align}
&  \frac{k  }{ 16 \pi^{ 2} } \mathcal{M}^{(2b, \mathbf{ k}_2)}_{L M, L' M'} (  k ) = \delta_{L M, L' M' }     \frac{ i k  }{ 16 \pi^{ 2} }      \nonumber \\
& + \delta_{L M, L' M' }      \int \frac{ p^{2}d  p }{(2\pi)^3}  \left (   \frac{  g^{(2b)}_{L}(p) }{g^{(2b)}_{L}(k)  }   \right)^2 \frac{1}{ k^2 -  p^2 }      \nonumber \\
& -       \frac{1}{L^3} \sum_{\mathbf{ p} = \frac{2\pi \mathbf{ n}}{L} +  \frac{ \mathbf{ k}_2}{2} , \mathbf{ n} \in \mathbb{Z}^3  }   \frac{ g^{(2b)}_{L} (p ) g^{(2b)}_{L'} (p )}{   g^{(2b)}_{L}(k) g^{(2b)}_{L'}(k)}   \frac{        Y^*_{L  M}  (\mathbf{ \hat{p}}) Y_{L' M'}  (\mathbf{ \hat{p}})    }{  k^2 -  \mathbf{ p}^2  }     ,
\end{align}
and
\begin{align}
 & \frac{ (mE)^{2} }{128 \pi^{ 5} }   \mathcal{M}^{(3b)}_{[J], [J']} (\sqrt{m E})   = \delta_{[J], [J']}  \frac{ i (mE)^{2} }{128 \pi^{ 5} }   \nonumber \\
& + \delta_{[J], [J']}      \int \frac{ P^{5}d  P }{(2\pi)^6 }   \left (\frac{g^{(3b)}_{J} ( P)}{ g^{(3b)}_{J} ( \sqrt{m E})} \right )^2       \frac{ 1  }{ mE - P^2}      \nonumber \\
& -       \frac{1}{L^6} \sum_{\mathbf{ p}_1,   \mathbf{ p}_{2}}   \frac{ g^{(3b)}_{J} ( P) g^{(3b)}_{J'} ( P) }{g^{(3b)}_{J} ( \sqrt{mE}) g^{(3b)}_{J'} ( \sqrt{mE})  } \frac{        Y^*_{[J]}  (\mathbf{ \hat{P}}) Y_{[J']}  (\mathbf{ \hat{P}})    }{ m E  - P^2}     .
\end{align}
  The two-body phase shift $ \delta^{(2b)}_{L}$ is defined in a conventional way, which may be modeled and constrained by experimental data. The unconventional three-body phase shift $\delta^{(3b)}_J$ may be interpreted as  scattering of one particle    off    a short-range potential in $6D$. It may only serve as a mathematically convenient tool for the modeling of  three-body resonance of total spin-$J$.

\subsection{Quantization condition with separable interactions approximation}
Algebra solutions of $\tau^{(2b)}$  in Eq.(\ref{tau2bsol})  and  $\tau^{(3b)}$ in  Eq.(\ref{tau3bsol})   suggest that partial expansion of  $\mathcal{T}^{(2b)} (\mathbf{ k}_1,  \mathbf{ k}_2) $  may have the form of
\begin{equation}
\mathcal{T}^{(2b)} (\mathbf{ k}_1,  \mathbf{ k}_2)  =  \sum_{L M } Y_{L M} ( \mathbf{\hat{ k}}_{13} )  g^{(2b)}_L (k_{13})\mathcal{T}^{(2b)}_{LM} (  \mathbf{ k}_2) .
\end{equation}
The separable form of $\mathcal{T}^{(2b)} (\mathbf{ k}_1,  \mathbf{ k}_2)$ thus allow one to further reduce Faddeev equations, Eq.(\ref{T2bFaddeev}) and Eq.(\ref{T3bFaddeev}), to 
\begin{widetext}
  \begin{align}
          \mathcal{T}^{(2b)}_{L M} ( \mathbf{ k}_2)  & =  -    \frac{1}{L^3}    \sum_{ 
              \mathbf{ p}_{2}    }  \sum_{L'  M' }    \frac{    2  \tau^{(2b)}_{L M} (  \mathbf{ p}_2+ \frac{\mathbf{ k}_2}{2})    g^{(2b)}_{L'} ( | \mathbf{ k}_{2}  + \frac{ \mathbf{ p}_2 }{2} |)    Y_{L' M'}  (  \mathbf{ k}_{2}  + \frac{ \mathbf{ p}_2 }{2} ) }{mE - \frac{3}{4} \mathbf{ k}^2_2 - (  \mathbf{ p}_{2}  + \frac{ \mathbf{ k}_2 }{2}  )^2   }                \mathcal{T}^{(2b)}_{L' M'} ( \mathbf{ p}_2)     \nonumber \\
              & +       \frac{1}{L^3}   \sum_{ 
              \mathbf{ p}_{2}    }   \sum_{L' M'}   \bigg [   \frac{3}{L^6}    \sum_{  \mathbf{ k}_1  ,
              \mathbf{ p}_1     }   \frac{    \tau^{(2 b)}_{L M} (  \mathbf{ k}_{13} )         }{mE  - \mathbf{ K}^2 }          \frac{\tau^{(3b)} (   \mathbf{ K}    ;\mathbf{ P} )   g^{(2b)}_{L'} (   p_{13})    Y_{L' M'}  ( \mathbf{ \hat{p}}_{13}     ) }{mE - \mathbf{ P}^2    }        \bigg ]  \mathcal{T}^{(2b)}_{L' M'} ( \mathbf{ p}_2)    ,
 \end{align}
  \end{widetext}
where $\tau^{(2b)}_{LM }(   \mathbf{ k}'_{13})  $ is  defined by relation
\begin{align}
 &\tau^{(2b)} (\mathbf{ k}_{13};  \mathbf{ k}'_{13})     =  \sum_{L M }     Y_{L M} ( \mathbf{\hat{ k}}_{13} )   g^{(2b)}_L (k_{13})    \tau^{(2b)}_{LM } (   \mathbf{ k}'_{13})     . 
\end{align}
 Therefore, a partially expanded  quantization condition is given by
   \begin{widetext}
  \begin{align}
         \det \bigg[   \delta_{LM, L'M'} L^3\delta_{ \mathbf{ k}_2, \mathbf{ p}_2}     & +   \frac{    2  \tau^{(2b)}_{L M} (  \mathbf{ p}_2+ \frac{\mathbf{ k}_2}{2})     g^{(2b)}_{L'} ( | \mathbf{ k}_{2}  + \frac{ \mathbf{ p}_2 }{2} |)    Y_{L' M'}  (  \mathbf{ k}_{2}  + \frac{ \mathbf{ p}_2 }{2} )        }{mE - \frac{3}{4} \mathbf{ k}^2_2 - (  \mathbf{ p}_{2}  + \frac{ \mathbf{ k}_2 }{2}  )^2   }     \nonumber \\
       &      -      \frac{1}{L^6}    \sum_{  \mathbf{ k}_1 , \mathbf{ p}_1      }        \frac{  3  \tau^{(2 b)}_{L M} (  \mathbf{ k}_{13}  )   \tau^{(3b)} (   \mathbf{ K}    ;\mathbf{ P} )   g^{(2b)}_{L'} (   p_{13})    Y_{L' M'}  ( \mathbf{ \hat{p}}_{13}     )   }{ \left (  mE  -  \mathbf{ K}^2 \right ) \left ( mE - \mathbf{ P}^2    \right )   }       \bigg ]   =0  .
 \end{align}
 \end{widetext}

As a specific example, let's consider a simple case with only $S$-wave contributions in both two-body and three-body channels,  that is to say, $J=L_{13}= L_{(13)2}=0$.  Thus  $\tau^{(2b)}$ and $\tau^{(3b)} $ are given respectively by   phase shifts $\delta^{(2b)}_{L_{13}=0}$ and $\delta^{(3b)}_{J=0}$ only,
\begin{align}
&  \tau^{(2b)} (\mathbf{ k}_{13};  \mathbf{ k}'_{13})   =   \tau^{(2b, \mathbf{ k}_2)}  ( \sqrt{mE - \frac{3}{4} \mathbf{ k}^2_2}  )  =  \frac{4\pi }{ \sqrt{mE - \frac{3}{4} \mathbf{ k}^2_2} }     \nonumber \\
&  \times \frac{1}{ \cot \delta^{(2b)}_0 (\sqrt{mE - \frac{3}{4} \mathbf{ k}^2_2} ) - \mathcal{M}^{(2b, \mathbf{ k}_2)}_{ 00, 00}  (\sqrt{mE - \frac{3}{4} \mathbf{ k}^2_2})  },  
\end{align}
and  
\begin{align}
&  \tau^{(3b)} (\mathbf{ K} ;   \mathbf{ K}'  ) = \tau^{(3b)} (\sqrt{mE})  \nonumber \\
& =   \frac{ 128 \pi^2}{ (mE)^{2} }  \frac{1}{   \cot \delta^{(3b)}_0 ( \sqrt{m E})  - \mathcal{M}^{(3b)}_{[0], [0]} (\sqrt{m E})   } .
\end{align} 
The quantization condition in this case is given by a simple form,
 \begin{align}
         \det & \bigg[    L^3\delta_{ \mathbf{ k}_2, \mathbf{ p}_2}     +   \frac{    2    \tau^{(2b, \mathbf{ k}_2)}  ( \sqrt{mE - \frac{3}{4} \mathbf{ k}^2_2}  )        }{mE - \frac{3}{4} \mathbf{ k}^2_2 - (  \mathbf{ p}_{2}  + \frac{ \mathbf{ k}_2 }{2}  )^2   }     \nonumber \\
       &      +       \frac{1}{L^6}    \sum_{  \mathbf{ k}_1 , \mathbf{ p}_1      }        \frac{  3    \tau^{(2b, \mathbf{ k}_2)}  ( \sqrt{mE - \frac{3}{4} \mathbf{ k}^2_2}  )   \tau^{(3b)} (\sqrt{mE})   }{ \left (  mE  -  \mathbf{ K}^2 \right ) \left ( mE - \mathbf{ P}^2    \right )   }       \bigg ]   =0  .
 \end{align}
 The two-body and three-body resonances hence can be inserted through modeling of $\delta^{(2b)}_{L_{13}=0}$ and $\delta^{(3b)}_{J=0}$.

\section{Summary}\label{summ}

In summary, with separable interaction approximation, we show   that the subprocess transition amplitudes are L\"uscher  formula-like, and   the quantization condition may be formulated in terms of both two-body and three-body phase shifts that may be useful for describing resonances in few-body interactions. Two-body phase shift may be modeled and constrained by experimental data, and three-body phase shift may serve as a convenient tool for inserting three-body resonances with a specific spin into quantization condition.

\begin{acknowledgements}
 We   acknowledge support from the Department of Physics and Engineering, California State University, Bakersfield, CA.   This research was supported in part by the National Science Foundation under Grant No. NSF PHY-1748958.    
 \end{acknowledgements}

\appendix

\section{L\"uscher  formula in $D$-dimensional space}\label{HHexpand}

\subsection{Scattering in $D$-dimensional space}
Let's start with $N$-body  Schr\"odinger equation in center of mass frame,
 \begin{equation}
\left [  m E   +  \sum_{j=1}^{N-1} \nabla_{ \bm{\xi}_j }^2 \right ]   \psi (\bm{\xi}; \mathbf{ K})   = m V (\xi)  \psi (\bm{\xi}; \mathbf{ K})   .
\end{equation}
the  relative coordinates of $N$-particle are given by
\begin{align}
&   \bm{\xi}_j =\sqrt{\frac{2 j}{j+1}}  \left  (   \frac{1}{j} \sum_{i=1}^j \mathbf{ x}_i  - \mathbf{ x}_{j+1}  \right ),   \nonumber \\
&   \mathbf{ q}_j =\sqrt{\frac{ j}{2(j+1)}}  \left  (   \frac{1}{j} \sum_{i=1}^j \mathbf{ k}_i  -  \mathbf{ k}_{j+1} \right ),   j=1,\cdots, N-1,   
 \end{align}
 where $\mathbf{ x}_i$ and $\mathbf{ k}_i$ stand for the coordinate and momentum of i-th particle respectively. $D=3(N-1)$ dimensional vector $(\bm{\xi}, \mathbf{ K})$ are defined by  relative coordinates and momenta of particles,
 \begin{align}
 \bm{\xi} &= \{ \bm{\xi}_1, \bm{\xi}_2 , \cdots,  \bm{\xi}_{N-1}\}, \ \  \xi  =| \bm{\xi} | = \sqrt{ \sum_{j=1}^{N-1}   \bm{\xi}_j^2 },  \nonumber \\
 \mathbf{ K}& = \{  \mathbf{ q}_1, \mathbf{ q}_2 , \cdots,  \mathbf{ q}_{N-1}\}, \ \  K = | \mathbf{ K}|= \sqrt{ \sum_{j=1}^{N-1}    \mathbf{ q}_j^2 }.
 \end{align}

$D$-dimensional Laplace operator has a separable  form between radial and orbital terms  \cite{Ripelle:1983hh,Ripelle:1993gr},
\begin{equation}
  \nabla^2_{D} = \sum_{j=1}^{N-1} \nabla_{ \bm{\xi}_j }^2 = \frac{1}{\xi^{D-1}} \frac{\partial}{\partial \xi}   \xi^{D-1}  \frac{ \partial}{\partial \xi} + \frac{\hat{L}^2 (  \Omega_D ) }{\xi^2},   
\end{equation}
where  $\hat{L}^2 (\Omega_D) $ is the grand orbital operator.   The eigen-states of orbital equation
\begin{equation}
 \hat{L}^2 (\Omega_D) Y_{[L]} (\Omega_D) = L(L + D -2)  Y_{[L]} (\Omega_D) 
\end{equation}
is given  by  hyperspherical harmonic  $Y_{[L]} (\Omega_D) $ \cite{Ripelle:1983hh,Ripelle:1993gr}, where $[L]$ is a set of $D-1$ quantum numbers, including total orbital angular momentum $L$.  Hyperspherical harmonic $Y_{[L]} (\Omega_D) $ basis define a complete set of orthonormal angular function in $D$-dimensional space,
\begin{equation}
\int d \Omega_D  Y^*_{[L]} (\Omega_D) Y_{[L']} (\Omega_D)= \delta_{[L], [L']}. 
\end{equation}

The  scattering in $D$-dimensional space can also be described by Lippmann-Schwinger equation
\begin{align}
 & \psi (\bm{\xi}; \mathbf{ K})   =  e^{i  \mathbf{ K} \cdot \bm{\xi}} +  \int d \bm{\xi}' G_D ( \bm{\xi} - \bm{\xi}' ; E) m V(\xi')   \psi (\bm{\xi}'; \mathbf{ K}),  \nonumber \\
 & G_D ( \bm{\xi} - \bm{\xi}' ; E) = \int \frac{ d \mathbf{ Q} }{(2\pi)^D} \frac{ e^{i  \mathbf{ Q} \cdot  (\bm{\xi} - \bm{\xi}' )}   }{ mE- \mathbf{ Q}^2}, \label{LSpsieq}
\end{align}
where Green's function satisfies equation
\begin{equation}
\left [  m E +  \nabla^2_{D}  \right ]  G_D ( \bm{\xi} - \bm{\xi}' ; E) =    \delta(  \bm{\xi} - \bm{\xi}' ) .
\end{equation}
The analytic expression of Green's function and its partial wave expansion in terms of hyperspherical harmonic basis are given respectively by
\begin{equation}
G_D ( \bm{\xi}  ;  E) = - \frac{i }{4}  \frac{ (mE)^{\frac{D}{2}-1}}{(2\pi)^{\frac{D}{2}-1}} \frac{ H^{(1)}_{\frac{D}{2}-1} (\sqrt{mE}  \xi )}{ ( \sqrt{mE}  \xi)^{\frac{D}{2}-1} },
\end{equation}
and
\begin{align}
& G_D ( \bm{\xi} - \bm{\xi}' ;  E)   \stackrel{ \xi > \xi'}{ =} - i (mE)^{\frac{D-2}{2}}  \nonumber \\
& \times  \sum_{[L]}  Y_{[L]} (\Omega_{\bm{\xi}})  \mathcal{H}^{(1)}_L ( \sqrt{m E} \xi)  \mathcal{J}_L (  \sqrt{m E} \xi') Y^*_{[L]} (\Omega_{\bm{\xi}'}), \label{G0HHexpand}
\end{align}
where
\begin{equation}
 \mathcal{J}_L(z) = \sqrt{\frac{\pi}{2}} \frac{ J_{L+ \frac{D-2}{2}} (z)}{z^{\frac{D}{2}-1}}, \ \  \mathcal{N}_L(z) = \sqrt{\frac{\pi}{2}} \frac{ N_{L+ \frac{D-2}{2}} (z)}{z^{\frac{D}{2}-1}} ,
\end{equation}
and
\begin{equation}
\mathcal{H}^{(1)}_L(z) =  \mathcal{J}_L(z)  + i  \mathcal{N}_L(z) .
\end{equation}
Assuming  potential $V(\xi)$ is spherical and short-range, and also using partial wave expansion of plane wave in $D$-dimensional space,
\begin{align}
 e^{i  \mathbf{ K} \cdot \bm{\xi}}  =  \sqrt{\frac{2}{\pi}} (2\pi)^{\frac{D}{2}} \sum_{[L]}  i^L Y_{[L]} (\Omega_{\bm{\xi}}) Y^*_{[L]} (\Omega_{\mathbf{ K}}) \mathcal{J}_L ( \sqrt{mE} \xi), \label{planeexpand}
\end{align}
 the asymptotic form of wave function is obtained,
\begin{align}
 & \psi (\bm{\xi}; \mathbf{ K})    \stackrel{Large \ \xi }{\rightarrow}  \sqrt{\frac{2}{\pi}} (2\pi)^{\frac{D}{2}} \sum_{[L]}  i^L Y_{[L]} (\Omega_{\bm{\xi}}) Y^*_{[L]} (\Omega_{\mathbf{ K}})   \nonumber \\
  & \times \left [ \mathcal{J}_L ( \sqrt{mE} \xi)  + i   f^{(D)}_L( \sqrt{mE} )  \mathcal{H}^{(1)}_L (\sqrt{mE} \xi)  \right ],
\end{align}
where   $ f^{(D)}_L$ is defined by
\begin{align}
&   \sqrt{\frac{2}{\pi}} \frac{ (2\pi)^{\frac{D}{2}} }{(mE)^{\frac{D-2}{2}}}  i^L  f^{(D)}_L( \sqrt{ m E}) Y^*_{[L]} (\Omega_{\mathbf{ K}})    \nonumber \\
& =  -    \int d \bm{\xi}'  Y^*_{[L]} (\Omega_{\bm{\xi}'})   \mathcal{J}_L (  \sqrt{mE} \xi')  m V(\xi')   \psi (\bm{\xi}' ; \mathbf{ K}). \label{fDLeq}
\end{align}
Thus  $ f^{(D)}_L$ may be interpreted as partial wave scattering amplitude in $D$-dimensional space, and it can be parameterized in terms of $D$-dimensional phase shift $\delta_L^{(D)} (k_E) $  \cite{Ripelle:1983hh,Ripelle:1993gr} by
\begin{equation}
 f^{(D)}_L( \sqrt{ mE}) = \frac{1}{\cot \delta_L^{(D)} (\sqrt{m E}) - i } .
\end{equation}

\subsection{Lippmann-Schwinger equation in momentum space and separable potential approximation}
The off-shell transition amplitude between initial and final momentum states $| \mathbf{ K} \rangle$  and $| \mathbf{ K}' \rangle$  may be introduced by
\begin{equation}
t^{(D)}( \mathbf{ K} , \mathbf{ K}' ) = - \int d \bm{ \xi }   e^{ -i  \mathbf{ K}' \cdot \bm{\xi}}  m V(\xi)   \psi (\bm{\xi}; \mathbf{ K}),
\end{equation}
thus Eq.(\ref{LSpsieq})  can be converted into momentum space Lippmann-Schwinger equation,
\begin{align}
& t^{(D)}( \mathbf{ K} , \mathbf{ K}' ) = - m  \widetilde{V} (| \mathbf{ K} - \mathbf{ K}'  | )  \nonumber \\
& + \int \frac{d \mathbf{ Q}}{(2\pi)^D} \frac{ m  \widetilde{V} ( | \mathbf{ K} - \mathbf{ Q} |)  }{ mE - \mathbf{ Q}^2} t^{(D)}( \mathbf{ Q} , \mathbf{ K}' )  .
\end{align}
The partial wave expansion of above equation yields
\begin{align}
& t^{(D)}_L( K, K' ) = - m  \widetilde{V}_L ( K, K' )  \nonumber \\
& + \int \frac{ Q^{D-1} d  Q}{(2\pi)^D} \frac{ m  \widetilde{V}_L (  K ,  Q)  }{ mE - \mathbf{ Q}^2} t^{(D)}_L ( Q, K' )  ,
\end{align}
where  the expansion relations of potential and amplitude are given by
\begin{equation}
\widetilde{V} (| \mathbf{ K} - \mathbf{ K}' | )  = \sum_{[L]}  Y_{[L]}  (\mathbf{ \hat{K}})   \widetilde{V}_L ( K, K' )  Y^*_{[L]}  (\mathbf{ \hat{K}}'),  
\end{equation}
and
\begin{equation}
t^{(D)}( \mathbf{ K} , \mathbf{ K}' )  =  \sum_{[L]}  Y_{[L]}  (\mathbf{ \hat{K}})   t_L ( K, K' ) Y^*_{[L]}  (\mathbf{ \hat{K}}') .
\end{equation}

Under assumption of separable potential, 
\begin{equation}
\widetilde{V}_L ( K, K' )  = g^{(D)}_L (K) V_L  g^{(D)}_L (K') , \label{seppot}
\end{equation}
where $g_L^{(D)}$ and $V_L$ stand for the form factor and interaction strength of potential, thus
a closed algebra form of off-shell partial wave amplitude, $t^{(D)}_L( K, K' ) $, may be obtained, see \cite{Lovelace:1964mq},
\begin{equation}
 t^{(D}_L (K,K')  = - \frac{g^{(D)}_L (K)   g^{(D)}_L (K' )}{ \frac{1}{m V_L}  -  \int \frac{ Q^{D-1}d  Q }{(2\pi)^D} \frac{  \left (g_L^{(D)}(Q) \right )^2 }{ mE - \mathbf{ Q}^2}   }.
\end{equation}
Compared with  on-shell scattering amplitude $ f^{(D)}_L(\sqrt{mE})$ in Eq.(\ref{fDLeq}), we find
\begin{align}
 & t^{(D}_L (K,K')  = \frac{g^{(D)}_L (K)   g^{(D)}_L (K' ) }{ \left ( g^{(D)}_L (\sqrt{mE}) \right )^2  }       \nonumber \\
  & \times \frac{2}{\pi}   \frac{(2\pi)^{ D} }{ (mE)^{\frac{D-2}{2}} }  \frac{1}{\cot \delta^{(D)}_L (\sqrt{mE}) - i }  , 
\end{align}
and also a useful relation
\begin{align}
  &  \frac{1}{m V_L}  =   \int \frac{ Q^{D-1}d  Q }{(2\pi)^D} \frac{  \left ( g^{(D)}_L(Q) \right )^2 }{ mE - \mathbf{ Q}^2}     \nonumber \\
   & +   \left ( g^{(D)}_L (\sqrt{mE})   \right )^2  \frac{\pi}{2}       \frac{ (mE)^{\frac{D-2}{2}} }{(2\pi)^{ D} }  \left [ i -   \cot \delta^{(D)}_L (\sqrt{mE})    \right ].  \label{VLeq}
\end{align}
Therefore  off-shell partial wave amplitude, $t^{(D)}_L( K, K' ) $ may be modeled in terms of on-shell physical quantity: phase shifts $\delta^{(D)}_L (\sqrt{mE})$.

We remark that the separable potential approximation is in fact based on the assumption of hyperspherical short-range interaction. The hyperspherical partial wave expansion of momentum space potential is given by 
\begin{align}
& \widetilde{V} (| \mathbf{ K} - \mathbf{ K}' | )  = \int d \bm{ \xi }   e^{ -i ( \mathbf{ K} - \mathbf{ K}' ) \cdot \bm{\xi}}   V(\xi)  \nonumber \\
& \propto \sum_{[L]}  Y_{[L]}  (\mathbf{ \hat{K}})   \int \xi^{D-1} d \xi  \mathcal{J}_L (K \xi) V(\xi) \mathcal{J}_L ( K' \xi)  Y^*_{[L]}  (\mathbf{ \hat{K}}').
\end{align}
For short-range potential, asymptotically one obtains   
\begin{equation}
 \int \xi^{D-1} d \xi  \mathcal{J}_L (K \xi) V(\xi) \mathcal{J}_L ( K' \xi)  \sim K^L  V_L  {K'}^L,
\end{equation}
which thus yield the expression in Eq.(\ref{seppot}).  The separable potential approximation may be useful  for the modeling of sharp few-body resonances that are predominantly generated by quark and gluon dynamics. Hence, the hadron-hadron interactions may be well approximated by a  short-range energy dependent interaction, the Breit-Wigner formula is a good example of such an approximation.

\subsection{L\"uscher  formula in $D$-dimensional space and separable potential approximation}
Scattering solution in finite volume may be described by inhomogeneous Lippmann-Schwinger equation, 
\begin{align}
& \tau^{(D)} ( \mathbf{ K} , \mathbf{ K}' ) = - m \widetilde{V} ( | \mathbf{ K} - \mathbf{ K}' | )  \nonumber \\
&+ \frac{1}{L^D} \sum_{\mathbf{ p}_1, \cdots, \mathbf{ p}_{N-1}} \frac{ m   \widetilde{V} (| \mathbf{ K} - \mathbf{ Q}| )  }{ mE - \mathbf{ Q}^2}  \tau^{(D)} ( \mathbf{ Q} , \mathbf{ K}' ) ,
\end{align}
where $ \mathbf{ p}_i \in \frac{2\pi \mathbf{ n}}{L}, \mathbf{ n} \in \mathbb{Z}^3$ and $\mathbf{ Q}^2= \frac{1}{2} \sum_{i=1}^N \mathbf{ p}^2_i$. Considering partial wave expansion  again,
\begin{equation}
  \tau^{(D)}  ( \mathbf{ K} , \mathbf{ K}' )   =  \sum_{[L], [L']}  Y_{[L]}  (\mathbf{ \hat{K}})   \tau^{(D)} _{[L] ,[L'] } ( K, K' ) Y^*_{[L']}  (\mathbf{ \hat{K}}'),  
\end{equation}
one finds
\begin{align}
&    \tau^{(D)}_{[L] ,[L'] } ( K, K' )  = -     \delta_{[L], [L']}   m \widetilde{V}_{L} (K, K' )    \nonumber \\
& + \sum_{ [l]}  \frac{1}{L^D} \sum_{\mathbf{ p}_1, \cdots, \mathbf{ p}_{N-1}} \frac{       m \widetilde{V}_L (K, Q )  Y^*_{[L]}  (\mathbf{ \hat{Q}}) Y_{[l]}  (\mathbf{ \hat{Q}})   }{ mE - \mathbf{ Q}^2}  \nonumber \\
& \quad   \times   \tau^{(D)}_{[l] ,[L'] } ( Q, K' )   .
\end{align}
Again,  the separable potential given in Eq.(\ref{seppot}) suggests that $\tau^{(D)}_{[L] ,[L'] }$ may have the separable form of
\begin{equation}
 \tau^{(D)}_{[L] ,[L'] } ( K, K' ) = g_L (K) C_{[L] ,[L'] } (E)  g_{L'} (K'),
 \end{equation}
 where $C_{[L] ,[L'] } (E)$ satisfies a matrix equation,
 \begin{align}
&   C_{[L] ,[L'] } (E)  = -     \delta_{[L], [L']}  mV_L   + \sum_{ [l]}  \frac{1}{L^D} \sum_{\mathbf{ p}_1, \cdots, \mathbf{ p}_{N-1}}   \nonumber \\
&   \quad   \times  g^{(D)}_L (Q)   g^{(D)}_l (Q)  \frac{        Y^*_{[L]}  (\mathbf{ \hat{Q}}) Y_{[l]}  (\mathbf{ \hat{Q}})  }{ mE - \mathbf{ Q}^2}      C_{[l] ,[L'] } (E)    .
\end{align}
Hence, a closed algebra form of solution of off-shell solution of finite volume amplitude,  $ \tau^{(D)}_{[L] ,[L'] }  $,  is obtained,
\begin{equation}
 \tau^{(D)}_{[L] ,[L'] } ( K, K' )  =  \frac{ g_L (K) g_{L'} (K')  }{g_L (\sqrt{mE})g_{L'} (\sqrt{mE})}  \left [ \mathcal{D}( \sqrt{m E}) \right ]^{-1} _{[L], [L']}    ,
\end{equation}
where
\begin{align}
& \mathcal{D}_{[L], [L']} ( \sqrt{mE})  = -  \frac{ \delta_{[L], [L']} }{ g_L(\sqrt{m E}) m V_L g_{L'}(\sqrt{m E}) }   \nonumber \\
&   +       \frac{1}{L^D} \sum_{\mathbf{ p}_1, \cdots, \mathbf{ p}_{N-1}} \frac{  g_L (Q ) g_{L'} (Q )   }{ g_L(\sqrt{m E})  g_{L'}(\sqrt{m E}) }\frac{        Y^*_{[L]}  (\mathbf{ \hat{Q}}) Y_{[L']}  (\mathbf{ \hat{Q}})    }{ mE - \mathbf{ Q}^2}     .
\end{align}
Using relation given in Eq.(\ref{VLeq}),  thus  $\mathcal{D}_{[L], [L']} $ is linked to generalized   L\"uscher  formula in $D$-dimensional space,
\begin{align}
&   \frac{2}{\pi}      \frac{(2\pi)^{ D} }{(mE)^{\frac{D-2}{2}} }  i^{L-L'}  \mathcal{D}_{[L], [L']} ( \sqrt{mE})  \nonumber \\
&  =   \delta_{[L], [L']} \cot \delta^{(D)}_L ( \sqrt{m E})  -\mathcal{M}_{[L], [L']} ( \sqrt{mE})     ,
\end{align}
where $\mathcal{M}_{[L], [L']} $ is  generalized    L\"uscher's  zeta function in $D$-dimensional space,
\begin{align}
& \frac{\pi}{2}      \frac{(mE)^{\frac{D-2}{2}} }{(2\pi)^{ D} }   \mathcal{M}_{[L], [L']} ( \sqrt{mE})   =   i \delta_{[L], [L']}      \frac{\pi}{2}      \frac{(mE)^{\frac{D-2}{2}} }{(2\pi)^{ D} }       \nonumber \\
&  -    \frac{1}{L^D} \sum_{\mathbf{ p}_1, \cdots, \mathbf{ p}_{N-1}} \frac{ i^{L-L'} g_L (Q ) g_{L'} (Q )   }{ g_L(\sqrt{m E})  g_{L'}(\sqrt{m E}) }  \frac{        Y^*_{[L]}  (\mathbf{ \hat{Q}}) Y_{[L']}  (\mathbf{ \hat{Q}})    }{ mE - \mathbf{ Q}^2}    \nonumber \\
&   + \delta_{[L], [L']}       \int \frac{ Q^{D-1}d  Q }{(2\pi)^D}  \left (\frac{g^{(D)}_L(Q)}{g^{(D)}_L(\sqrt{mE})} \right )^2 \frac{ 1 }{ mE- \mathbf{ Q}^2}      . \label{zetafunction}
\end{align}
Therefore, the inverse of  $ \tau^{(D)}_{[L] ,[L'] }  $ is explicitly related to L\"uscher  formula  by
\begin{align}
&  \frac{2}{\pi}      \frac{(2\pi)^{ D} }{(mE)^{\frac{D-2}{2}} }  \left [\frac{g_L (\sqrt{mE})g_{L'} (\sqrt{mE})}{ g_L (K) g_{L'} (K')  } \tau^{(D)} ( K, K' ) \right ]^{-1}_{[L] ,[L'] }  \nonumber \\
&  = i^{L'-L} \left [   \delta_{[L], [L']} \cot \delta^{(D)}_L ( \sqrt{m E})  -\mathcal{M}_{[L], [L']} ( \sqrt{mE})  \right ]  .
\end{align}

Generalized  L\"uscher  zeta function can also be derived by considering hyperspherical harmonic basis function expansion of Green's function. In infinite volume,  the hyperspherical harmonic basis expansion of Green's function is given by,  
\begin{align}
&  \int \frac{ d \mathbf{ Q} }{(2\pi)^D} \frac{ e^{i  \mathbf{ Q} \cdot  (\bm{\xi} - \bm{\xi}' )}   }{ mE- \mathbf{ Q}^2}  \stackrel{ \xi > \xi'}{ =} - i (mE)^{\frac{D-2}{2}}  \nonumber \\
& \times  \sum_{[L]}  Y_{[L]} (\Omega_{\bm{\xi}})  \mathcal{H}^{(1)}_L ( \sqrt{m E} \xi)  \mathcal{J}_L (  \sqrt{m E} \xi') Y^*_{[L]} (\Omega_{\bm{\xi}'}). \label{G0expand}
\end{align}
Similarly to expansion of infinite volume Green's function, the expansion of finite volume Green's function may be written as,
\begin{align}
&  \frac{1}{L^D} \sum_{\mathbf{ p}_1, \cdots, \mathbf{ p}_{N-1}} \frac{  e^{i  \mathbf{ Q} \cdot  (\bm{\xi} - \bm{\xi}' )}     }{ mE - \mathbf{ Q}^2}    \stackrel{ \xi > \xi'}{ =}  (mE)^{\frac{D-2}{2}}  \sum_{[L] , [L']}   Y_{[L]} (\Omega_{\bm{\xi}})     \nonumber \\
& \times  \left [ \delta_{[L], [L']}     \mathcal{N}_L ( \sqrt{m E} \xi)  -  \mathcal{M}_{[L], [L']} ( \sqrt{mE})  \mathcal{J}_L ( \sqrt{m E} \xi)   \right ]\nonumber \\
& \times\mathcal{J}_{L'} (  \sqrt{m E} \xi') Y^*_{[L']} (\Omega_{\bm{\xi}'}). \label{GLexpand}
\end{align}
Combining Eq.(\ref{G0expand}) and Eq.(\ref{GLexpand}), we   obtain
\begin{align}
&  \frac{1}{L^D} \sum_{\mathbf{ p}_1, \cdots, \mathbf{ p}_{N-1}} \frac{  e^{i  \mathbf{ Q} \cdot  (\bm{\xi} - \bm{\xi}' )}     }{ mE - \mathbf{ Q}^2}   - \int \frac{ d \mathbf{ Q} }{(2\pi)^D} \frac{ e^{i  \mathbf{ Q} \cdot  (\bm{\xi} - \bm{\xi}' )}   }{ mE- \mathbf{ Q}^2} \nonumber \\
&   \stackrel{ \xi > \xi'}{ =}  (mE)^{\frac{D-2}{2}}  \sum_{[L] , [L']}   Y_{[L]} (\Omega_{\bm{\xi}})     \mathcal{J}_L ( \sqrt{m E} \xi)   \nonumber \\
& \times  \left [ i \delta_{[L], [L']}       -  \mathcal{M}_{[L], [L']} ( \sqrt{mE})    \right  ] \mathcal{J}_{L'} (  \sqrt{m E} \xi') Y^*_{[L']} (\Omega_{\bm{\xi}'}).  
\end{align}
Next, using plane wave expansion formula given in Eq.(\ref{planeexpand}) and also replacing $g^{(D)}_L (k)$   by $k^L$, we thus find again Eq.(\ref{zetafunction}), which may also suggest that the form factor, $g^{(D)}_L $, may be chosen as $g^{(D)}_L (k)  \sim k^L$.

\bibliography{ALL-REF.bib}

\begin{thebibliography}{57}%
\makeatletter
\providecommand \@ifxundefined [1]{%
 \@ifx{#1\undefined}
}%
\providecommand \@ifnum [1]{%
 \ifnum #1\expandafter \@firstoftwo
 \else \expandafter \@secondoftwo
 \fi
}%
\providecommand \@ifx [1]{%
 \ifx #1\expandafter \@firstoftwo
 \else \expandafter \@secondoftwo
 \fi
}%
\providecommand \natexlab [1]{#1}%
\providecommand \enquote  [1]{``#1''}%
\providecommand \bibnamefont  [1]{#1}%
\providecommand \bibfnamefont [1]{#1}%
\providecommand \citenamefont [1]{#1}%
\providecommand \href@noop [0]{\@secondoftwo}%
\providecommand \href [0]{\begingroup \@sanitize@url \@href}%
\providecommand \@href[1]{\@@startlink{#1}\@@href}%
\providecommand \@@href[1]{\endgroup#1\@@endlink}%
\providecommand \@sanitize@url [0]{\catcode `\\12\catcode `\$12\catcode
  `\&12\catcode `\#12\catcode `\^12\catcode `\_12\catcode `\%12\relax}%
\providecommand \@@startlink[1]{}%
\providecommand \@@endlink[0]{}%
\providecommand \url  [0]{\begingroup\@sanitize@url \@url }%
\providecommand \@url [1]{\endgroup\@href {#1}{\urlprefix }}%
\providecommand \urlprefix  [0]{URL }%
\providecommand \Eprint [0]{\href }%
\providecommand \doibase [0]{http://dx.doi.org/}%
\providecommand \selectlanguage [0]{\@gobble}%
\providecommand \bibinfo  [0]{\@secondoftwo}%
\providecommand \bibfield  [0]{\@secondoftwo}%
\providecommand \translation [1]{[#1]}%
\providecommand \BibitemOpen [0]{}%
\providecommand \bibitemStop [0]{}%
\providecommand \bibitemNoStop [0]{.\EOS\space}%
\providecommand \EOS [0]{\spacefactor3000\relax}%
\providecommand \BibitemShut  [1]{\csname bibitem#1\endcsname}%
\let\auto@bib@innerbib\@empty
%</preamble>
\bibitem [{\citenamefont {Kambor}\ \emph {et~al.}(1996)\citenamefont {Kambor},
  \citenamefont {Wiesendanger},\ and\ \citenamefont {Wyler}}]{Kambor:1995yc}%
  \BibitemOpen
  \bibfield  {author} {\bibinfo {author} {\bibfnamefont {J.}~\bibnamefont
  {Kambor}}, \bibinfo {author} {\bibfnamefont {C.}~\bibnamefont
  {Wiesendanger}}, \ and\ \bibinfo {author} {\bibfnamefont {D.}~\bibnamefont
  {Wyler}},\ }\href {\doibase 10.1016/0550-3213(95)00676-1} {\bibfield
  {journal} {\bibinfo  {journal} {Nucl. Phys.}\ }\textbf {\bibinfo {volume}
  {B465}},\ \bibinfo {pages} {215} (\bibinfo {year} {1996})},\ \Eprint
  {http://arxiv.org/abs/hep-ph/9509374} {arXiv:hep-ph/9509374 [hep-ph]}
  \BibitemShut {NoStop}%
%%CITATION = HEP-PH/9509374;%%
\bibitem [{\citenamefont {Anisovich}\ and\ \citenamefont
  {Leutwyler}(1996)}]{Anisovich:1996tx}%
  \BibitemOpen
  \bibfield  {author} {\bibinfo {author} {\bibfnamefont {A.~V.}\ \bibnamefont
  {Anisovich}}\ and\ \bibinfo {author} {\bibfnamefont {H.}~\bibnamefont
  {Leutwyler}},\ }\href {\doibase 10.1016/0370-2693(96)00192-X} {\bibfield
  {journal} {\bibinfo  {journal} {Phys. Lett.}\ }\textbf {\bibinfo {volume}
  {B375}},\ \bibinfo {pages} {335} (\bibinfo {year} {1996})},\ \Eprint
  {http://arxiv.org/abs/hep-ph/9601237} {arXiv:hep-ph/9601237 [hep-ph]}
  \BibitemShut {NoStop}%
%%CITATION = HEP-PH/9601237;%%
\bibitem [{\citenamefont {Schneider}\ \emph {et~al.}(2011)\citenamefont
  {Schneider}, \citenamefont {Kubis},\ and\ \citenamefont
  {Ditsche}}]{Schneider:2010hs}%
  \BibitemOpen
  \bibfield  {author} {\bibinfo {author} {\bibfnamefont {S.~P.}\ \bibnamefont
  {Schneider}}, \bibinfo {author} {\bibfnamefont {B.}~\bibnamefont {Kubis}}, \
  and\ \bibinfo {author} {\bibfnamefont {C.}~\bibnamefont {Ditsche}},\ }\href
  {\doibase 10.1007/JHEP02(2011)028} {\bibfield  {journal} {\bibinfo  {journal}
  {JHEP}\ }\textbf {\bibinfo {volume} {02}},\ \bibinfo {pages} {028} (\bibinfo
  {year} {2011})},\ \Eprint {http://arxiv.org/abs/1010.3946} {arXiv:1010.3946
  [hep-ph]} \BibitemShut {NoStop}%
%%CITATION = ARXIV:1010.3946;%%
\bibitem [{\citenamefont {Kampf}\ \emph {et~al.}(2011)\citenamefont {Kampf},
  \citenamefont {Knecht}, \citenamefont {Novotny},\ and\ \citenamefont
  {Zdrahal}}]{Kampf:2011wr}%
  \BibitemOpen
  \bibfield  {author} {\bibinfo {author} {\bibfnamefont {K.}~\bibnamefont
  {Kampf}}, \bibinfo {author} {\bibfnamefont {M.}~\bibnamefont {Knecht}},
  \bibinfo {author} {\bibfnamefont {J.}~\bibnamefont {Novotny}}, \ and\
  \bibinfo {author} {\bibfnamefont {M.}~\bibnamefont {Zdrahal}},\ }\href
  {\doibase 10.1103/PhysRevD.84.114015} {\bibfield  {journal} {\bibinfo
  {journal} {Phys. Rev.}\ }\textbf {\bibinfo {volume} {D84}},\ \bibinfo {pages}
  {114015} (\bibinfo {year} {2011})},\ \Eprint {http://arxiv.org/abs/1103.0982}
  {arXiv:1103.0982 [hep-ph]} \BibitemShut {NoStop}%
%%CITATION = ARXIV:1103.0982;%%
\bibitem [{\citenamefont {Guo}\ \emph {et~al.}(2015)\citenamefont {Guo},
  \citenamefont {Danilkin}, \citenamefont {Schott}, \citenamefont
  {Fernández-Ramírez}, \citenamefont {Mathieu},\ and\ \citenamefont
  {Szczepaniak}}]{Guo:2015zqa}%
  \BibitemOpen
  \bibfield  {author} {\bibinfo {author} {\bibfnamefont {P.}~\bibnamefont
  {Guo}}, \bibinfo {author} {\bibfnamefont {I.~V.}\ \bibnamefont {Danilkin}},
  \bibinfo {author} {\bibfnamefont {D.}~\bibnamefont {Schott}}, \bibinfo
  {author} {\bibfnamefont {C.}~\bibnamefont {Fernández-Ramírez}}, \bibinfo
  {author} {\bibfnamefont {V.}~\bibnamefont {Mathieu}}, \ and\ \bibinfo
  {author} {\bibfnamefont {A.~P.}\ \bibnamefont {Szczepaniak}},\ }\href
  {\doibase 10.1103/PhysRevD.92.054016} {\bibfield  {journal} {\bibinfo
  {journal} {Phys. Rev.}\ }\textbf {\bibinfo {volume} {D92}},\ \bibinfo {pages}
  {054016} (\bibinfo {year} {2015})},\ \Eprint
  {http://arxiv.org/abs/1505.01715} {arXiv:1505.01715 [hep-ph]} \BibitemShut
  {NoStop}%
%%CITATION = ARXIV:1505.01715;%%
\bibitem [{\citenamefont {Guo}\ \emph {et~al.}(2017)\citenamefont {Guo},
  \citenamefont {Danilkin}, \citenamefont {Fernández-Ramírez}, \citenamefont
  {Mathieu},\ and\ \citenamefont {Szczepaniak}}]{Guo:2016wsi}%
  \BibitemOpen
  \bibfield  {author} {\bibinfo {author} {\bibfnamefont {P.}~\bibnamefont
  {Guo}}, \bibinfo {author} {\bibfnamefont {I.~V.}\ \bibnamefont {Danilkin}},
  \bibinfo {author} {\bibfnamefont {C.}~\bibnamefont {Fernández-Ramírez}},
  \bibinfo {author} {\bibfnamefont {V.}~\bibnamefont {Mathieu}}, \ and\
  \bibinfo {author} {\bibfnamefont {A.~P.}\ \bibnamefont {Szczepaniak}},\
  }\href {\doibase 10.1016/j.physletb.2017.05.092} {\bibfield  {journal}
  {\bibinfo  {journal} {Phys. Lett.}\ }\textbf {\bibinfo {volume} {B771}},\
  \bibinfo {pages} {497} (\bibinfo {year} {2017})},\ \Eprint
  {http://arxiv.org/abs/1608.01447} {arXiv:1608.01447 [hep-ph]} \BibitemShut
  {NoStop}%
%%CITATION = ARXIV:1608.01447;%%
\bibitem [{\citenamefont {Colangelo}\ \emph {et~al.}(2017)\citenamefont
  {Colangelo}, \citenamefont {Lanz}, \citenamefont {Leutwyler},\ and\
  \citenamefont {Passemar}}]{Colangelo:2016jmc}%
  \BibitemOpen
  \bibfield  {author} {\bibinfo {author} {\bibfnamefont {G.}~\bibnamefont
  {Colangelo}}, \bibinfo {author} {\bibfnamefont {S.}~\bibnamefont {Lanz}},
  \bibinfo {author} {\bibfnamefont {H.}~\bibnamefont {Leutwyler}}, \ and\
  \bibinfo {author} {\bibfnamefont {E.}~\bibnamefont {Passemar}},\ }\href
  {\doibase 10.1103/PhysRevLett.118.022001} {\bibfield  {journal} {\bibinfo
  {journal} {Phys. Rev. Lett.}\ }\textbf {\bibinfo {volume} {118}},\ \bibinfo
  {pages} {022001} (\bibinfo {year} {2017})},\ \Eprint
  {http://arxiv.org/abs/1610.03494} {arXiv:1610.03494 [hep-ph]} \BibitemShut
  {NoStop}%
%%CITATION = ARXIV:1610.03494;%%
\bibitem [{\citenamefont {Efimov}(1970)}]{Efimov:1970zz}%
  \BibitemOpen
  \bibfield  {author} {\bibinfo {author} {\bibfnamefont {V.}~\bibnamefont
  {Efimov}},\ }\href {\doibase 10.1016/0370-2693(70)90349-7} {\bibfield
  {journal} {\bibinfo  {journal} {Phys. Lett.}\ }\textbf {\bibinfo {volume}
  {33B}},\ \bibinfo {pages} {563} (\bibinfo {year} {1970})}\BibitemShut
  {NoStop}%
%%CITATION = PHLTA,33B,563;%%
\bibitem [{\citenamefont {Braaten}\ and\ \citenamefont
  {Hammer}(2006)}]{Braaten:2004rn}%
  \BibitemOpen
  \bibfield  {author} {\bibinfo {author} {\bibfnamefont {E.}~\bibnamefont
  {Braaten}}\ and\ \bibinfo {author} {\bibfnamefont {H.~W.}\ \bibnamefont
  {Hammer}},\ }\href {\doibase 10.1016/j.physrep.2006.03.001} {\bibfield
  {journal} {\bibinfo  {journal} {Phys. Rept.}\ }\textbf {\bibinfo {volume}
  {428}},\ \bibinfo {pages} {259} (\bibinfo {year} {2006})},\ \Eprint
  {http://arxiv.org/abs/cond-mat/0410417} {arXiv:cond-mat/0410417 [cond-mat]}
  \BibitemShut {NoStop}%
%%CITATION = COND-MAT/0410417;%%
\bibitem [{\citenamefont {Zhukov}\ \emph {et~al.}(1993)\citenamefont {Zhukov},
  \citenamefont {Danilin}, \citenamefont {Fedorov}, \citenamefont {Bang},
  \citenamefont {Thompson},\ and\ \citenamefont {Vaagen}}]{Zhukov:1993aw}%
  \BibitemOpen
  \bibfield  {author} {\bibinfo {author} {\bibfnamefont {M.~V.}\ \bibnamefont
  {Zhukov}}, \bibinfo {author} {\bibfnamefont {B.~V.}\ \bibnamefont {Danilin}},
  \bibinfo {author} {\bibfnamefont {D.~V.}\ \bibnamefont {Fedorov}}, \bibinfo
  {author} {\bibfnamefont {J.~M.}\ \bibnamefont {Bang}}, \bibinfo {author}
  {\bibfnamefont {I.~J.}\ \bibnamefont {Thompson}}, \ and\ \bibinfo {author}
  {\bibfnamefont {J.~S.}\ \bibnamefont {Vaagen}},\ }\href {\doibase
  10.1016/0370-1573(93)90141-Y} {\bibfield  {journal} {\bibinfo  {journal}
  {Phys. Rept.}\ }\textbf {\bibinfo {volume} {231}},\ \bibinfo {pages} {151}
  (\bibinfo {year} {1993})}\BibitemShut {NoStop}%
%%CITATION = PRPLC,231,151;%%
\bibitem [{\citenamefont {Hammer}\ \emph
  {et~al.}(2017{\natexlab{a}})\citenamefont {Hammer}, \citenamefont {Ji},\ and\
  \citenamefont {Phillips}}]{Hammer:2017tjm}%
  \BibitemOpen
  \bibfield  {author} {\bibinfo {author} {\bibfnamefont {H.~W.}\ \bibnamefont
  {Hammer}}, \bibinfo {author} {\bibfnamefont {C.}~\bibnamefont {Ji}}, \ and\
  \bibinfo {author} {\bibfnamefont {D.~R.}\ \bibnamefont {Phillips}},\ }\href
  {\doibase 10.1088/1361-6471/aa83db} {\bibfield  {journal} {\bibinfo
  {journal} {J. Phys.}\ }\textbf {\bibinfo {volume} {G44}},\ \bibinfo {pages}
  {103002} (\bibinfo {year} {2017}{\natexlab{a}})},\ \Eprint
  {http://arxiv.org/abs/1702.08605} {arXiv:1702.08605 [nucl-th]} \BibitemShut
  {NoStop}%
%%CITATION = ARXIV:1702.08605;%%
\bibitem [{\citenamefont {L{\"u}scher}(1991)}]{Luscher:1990ux}%
  \BibitemOpen
  \bibfield  {author} {\bibinfo {author} {\bibfnamefont {M.}~\bibnamefont
  {L{\"u}scher}},\ }\href {\doibase 10.1016/0550-3213(91)90366-6} {\bibfield
  {journal} {\bibinfo  {journal} {Nucl. Phys.}\ }\textbf {\bibinfo {volume}
  {B354}},\ \bibinfo {pages} {531} (\bibinfo {year} {1991})}\BibitemShut
  {NoStop}%
%%CITATION = NUPHA,B354,531;%%
\bibitem [{\citenamefont {Rummukainen}\ and\ \citenamefont
  {Gottlieb}(1995)}]{Rummukainen:1995vs}%
  \BibitemOpen
  \bibfield  {author} {\bibinfo {author} {\bibfnamefont {K.}~\bibnamefont
  {Rummukainen}}\ and\ \bibinfo {author} {\bibfnamefont {S.~A.}\ \bibnamefont
  {Gottlieb}},\ }\href {\doibase 10.1016/0550-3213(95)00313-H} {\bibfield
  {journal} {\bibinfo  {journal} {Nucl. Phys.}\ }\textbf {\bibinfo {volume}
  {B450}},\ \bibinfo {pages} {397} (\bibinfo {year} {1995})},\ \Eprint
  {http://arxiv.org/abs/hep-lat/9503028} {arXiv:hep-lat/9503028 [hep-lat]}
  \BibitemShut {NoStop}%
%%CITATION = HEP-LAT/9503028;%%
\bibitem [{\citenamefont {Christ}\ \emph {et~al.}(2005)\citenamefont {Christ},
  \citenamefont {Kim},\ and\ \citenamefont {Yamazaki}}]{Christ:2005gi}%
  \BibitemOpen
  \bibfield  {author} {\bibinfo {author} {\bibfnamefont {N.~H.}\ \bibnamefont
  {Christ}}, \bibinfo {author} {\bibfnamefont {C.}~\bibnamefont {Kim}}, \ and\
  \bibinfo {author} {\bibfnamefont {T.}~\bibnamefont {Yamazaki}},\ }\href
  {\doibase 10.1103/PhysRevD.72.114506} {\bibfield  {journal} {\bibinfo
  {journal} {Phys. Rev.}\ }\textbf {\bibinfo {volume} {D72}},\ \bibinfo {pages}
  {114506} (\bibinfo {year} {2005})},\ \Eprint
  {http://arxiv.org/abs/hep-lat/0507009} {arXiv:hep-lat/0507009 [hep-lat]}
  \BibitemShut {NoStop}%
%%CITATION = HEP-LAT/0507009;%%
\bibitem [{\citenamefont {Bernard}\ \emph {et~al.}(2008)\citenamefont
  {Bernard}, \citenamefont {Lage}, \citenamefont {Mei{\ss}ner},\ and\
  \citenamefont {Rusetsky}}]{Bernard:2008ax}%
  \BibitemOpen
  \bibfield  {author} {\bibinfo {author} {\bibfnamefont {V.}~\bibnamefont
  {Bernard}}, \bibinfo {author} {\bibfnamefont {M.}~\bibnamefont {Lage}},
  \bibinfo {author} {\bibfnamefont {U.-G.}\ \bibnamefont {Mei{\ss}ner}}, \ and\
  \bibinfo {author} {\bibfnamefont {A.}~\bibnamefont {Rusetsky}},\ }\href
  {\doibase 10.1088/1126-6708/2008/08/024} {\bibfield  {journal} {\bibinfo
  {journal} {JHEP}\ }\textbf {\bibinfo {volume} {08}},\ \bibinfo {pages} {024}
  (\bibinfo {year} {2008})},\ \Eprint {http://arxiv.org/abs/0806.4495}
  {arXiv:0806.4495 [hep-lat]} \BibitemShut {NoStop}%
%%CITATION = ARXIV:0806.4495;%%
\bibitem [{\citenamefont {He}\ \emph {et~al.}(2005)\citenamefont {He},
  \citenamefont {Feng},\ and\ \citenamefont {Liu}}]{He:2005ey}%
  \BibitemOpen
  \bibfield  {author} {\bibinfo {author} {\bibfnamefont {S.}~\bibnamefont
  {He}}, \bibinfo {author} {\bibfnamefont {X.}~\bibnamefont {Feng}}, \ and\
  \bibinfo {author} {\bibfnamefont {C.}~\bibnamefont {Liu}},\ }\href {\doibase
  10.1088/1126-6708/2005/07/011} {\bibfield  {journal} {\bibinfo  {journal}
  {JHEP}\ }\textbf {\bibinfo {volume} {07}},\ \bibinfo {pages} {011} (\bibinfo
  {year} {2005})},\ \Eprint {http://arxiv.org/abs/hep-lat/0504019}
  {arXiv:hep-lat/0504019 [hep-lat]} \BibitemShut {NoStop}%
%%CITATION = HEP-LAT/0504019;%%
\bibitem [{\citenamefont {Lage}\ \emph {et~al.}(2009)\citenamefont {Lage},
  \citenamefont {Mei{\ss}ner},\ and\ \citenamefont {Rusetsky}}]{Lage:2009zv}%
  \BibitemOpen
  \bibfield  {author} {\bibinfo {author} {\bibfnamefont {M.}~\bibnamefont
  {Lage}}, \bibinfo {author} {\bibfnamefont {U.-G.}\ \bibnamefont
  {Mei{\ss}ner}}, \ and\ \bibinfo {author} {\bibfnamefont {A.}~\bibnamefont
  {Rusetsky}},\ }\href {\doibase 10.1016/j.physletb.2009.10.055} {\bibfield
  {journal} {\bibinfo  {journal} {Phys. Lett.}\ }\textbf {\bibinfo {volume}
  {B681}},\ \bibinfo {pages} {439} (\bibinfo {year} {2009})},\ \Eprint
  {http://arxiv.org/abs/0905.0069} {arXiv:0905.0069 [hep-lat]} \BibitemShut
  {NoStop}%
%%CITATION = ARXIV:0905.0069;%%
\bibitem [{\citenamefont {D{\"o}ring}\ \emph {et~al.}(2011)\citenamefont
  {D{\"o}ring}, \citenamefont {Mei{\ss}ner}, \citenamefont {Oset},\ and\
  \citenamefont {Rusetsky}}]{Doring:2011vk}%
  \BibitemOpen
  \bibfield  {author} {\bibinfo {author} {\bibfnamefont {M.}~\bibnamefont
  {D{\"o}ring}}, \bibinfo {author} {\bibfnamefont {U.-G.}\ \bibnamefont
  {Mei{\ss}ner}}, \bibinfo {author} {\bibfnamefont {E.}~\bibnamefont {Oset}}, \
  and\ \bibinfo {author} {\bibfnamefont {A.}~\bibnamefont {Rusetsky}},\ }\href
  {\doibase 10.1140/epja/i2011-11139-7} {\bibfield  {journal} {\bibinfo
  {journal} {Eur. Phys. J.}\ }\textbf {\bibinfo {volume} {A47}},\ \bibinfo
  {pages} {139} (\bibinfo {year} {2011})},\ \Eprint
  {http://arxiv.org/abs/1107.3988} {arXiv:1107.3988 [hep-lat]} \BibitemShut
  {NoStop}%
%%CITATION = ARXIV:1107.3988;%%
\bibitem [{\citenamefont {Brice{\~n}o}\ and\ \citenamefont
  {Davoudi}(2013{\natexlab{a}})}]{Briceno:2012yi}%
  \BibitemOpen
  \bibfield  {author} {\bibinfo {author} {\bibfnamefont {R.~A.}\ \bibnamefont
  {Brice{\~n}o}}\ and\ \bibinfo {author} {\bibfnamefont {Z.}~\bibnamefont
  {Davoudi}},\ }\href {\doibase 10.1103/PhysRevD.88.094507} {\bibfield
  {journal} {\bibinfo  {journal} {Phys. Rev.}\ }\textbf {\bibinfo {volume}
  {D88}},\ \bibinfo {pages} {094507} (\bibinfo {year} {2013}{\natexlab{a}})},\
  \Eprint {http://arxiv.org/abs/1204.1110} {arXiv:1204.1110 [hep-lat]}
  \BibitemShut {NoStop}%
%%CITATION = ARXIV:1204.1110;%%
\bibitem [{\citenamefont {Hansen}\ and\ \citenamefont
  {Sharpe}(2012)}]{Hansen:2012tf}%
  \BibitemOpen
  \bibfield  {author} {\bibinfo {author} {\bibfnamefont {M.~T.}\ \bibnamefont
  {Hansen}}\ and\ \bibinfo {author} {\bibfnamefont {S.~R.}\ \bibnamefont
  {Sharpe}},\ }\href {\doibase 10.1103/PhysRevD.86.016007} {\bibfield
  {journal} {\bibinfo  {journal} {Phys. Rev.}\ }\textbf {\bibinfo {volume}
  {D86}},\ \bibinfo {pages} {016007} (\bibinfo {year} {2012})},\ \Eprint
  {http://arxiv.org/abs/1204.0826} {arXiv:1204.0826 [hep-lat]} \BibitemShut
  {NoStop}%
%%CITATION = ARXIV:1204.0826;%%
\bibitem [{\citenamefont {Guo}\ \emph {et~al.}(2013)\citenamefont {Guo},
  \citenamefont {Dudek}, \citenamefont {Edwards},\ and\ \citenamefont
  {Szczepaniak}}]{Guo:2012hv}%
  \BibitemOpen
  \bibfield  {author} {\bibinfo {author} {\bibfnamefont {P.}~\bibnamefont
  {Guo}}, \bibinfo {author} {\bibfnamefont {J.}~\bibnamefont {Dudek}}, \bibinfo
  {author} {\bibfnamefont {R.}~\bibnamefont {Edwards}}, \ and\ \bibinfo
  {author} {\bibfnamefont {A.~P.}\ \bibnamefont {Szczepaniak}},\ }\href
  {\doibase 10.1103/PhysRevD.88.014501} {\bibfield  {journal} {\bibinfo
  {journal} {Phys. Rev.}\ }\textbf {\bibinfo {volume} {D88}},\ \bibinfo {pages}
  {014501} (\bibinfo {year} {2013})},\ \Eprint {http://arxiv.org/abs/1211.0929}
  {arXiv:1211.0929 [hep-lat]} \BibitemShut {NoStop}%
%%CITATION = ARXIV:1211.0929;%%
\bibitem [{\citenamefont {Guo}(2013)}]{Guo:2013vsa}%
  \BibitemOpen
  \bibfield  {author} {\bibinfo {author} {\bibfnamefont {P.}~\bibnamefont
  {Guo}},\ }\href {\doibase 10.1103/PhysRevD.88.014507} {\bibfield  {journal}
  {\bibinfo  {journal} {Phys. Rev.}\ }\textbf {\bibinfo {volume} {D88}},\
  \bibinfo {pages} {014507} (\bibinfo {year} {2013})},\ \Eprint
  {http://arxiv.org/abs/1304.7812} {arXiv:1304.7812 [hep-lat]} \BibitemShut
  {NoStop}%
%%CITATION = ARXIV:1304.7812;%%
\bibitem [{\citenamefont {Kreuzer}\ and\ \citenamefont
  {Hammer}(2009)}]{Kreuzer:2008bi}%
  \BibitemOpen
  \bibfield  {author} {\bibinfo {author} {\bibfnamefont {S.}~\bibnamefont
  {Kreuzer}}\ and\ \bibinfo {author} {\bibfnamefont {H.~W.}\ \bibnamefont
  {Hammer}},\ }\href {\doibase 10.1016/j.physletb.2009.02.035} {\bibfield
  {journal} {\bibinfo  {journal} {Phys. Lett.}\ }\textbf {\bibinfo {volume}
  {B673}},\ \bibinfo {pages} {260} (\bibinfo {year} {2009})},\ \Eprint
  {http://arxiv.org/abs/0811.0159} {arXiv:0811.0159 [nucl-th]} \BibitemShut
  {NoStop}%
%%CITATION = ARXIV:0811.0159;%%
\bibitem [{\citenamefont {Kreuzer}\ and\ \citenamefont
  {Hammer}(2010)}]{Kreuzer:2009jp}%
  \BibitemOpen
  \bibfield  {author} {\bibinfo {author} {\bibfnamefont {S.}~\bibnamefont
  {Kreuzer}}\ and\ \bibinfo {author} {\bibfnamefont {H.~W.}\ \bibnamefont
  {Hammer}},\ }\href {\doibase 10.1140/epja/i2010-10910-6} {\bibfield
  {journal} {\bibinfo  {journal} {Eur. Phys. J.}\ }\textbf {\bibinfo {volume}
  {A43}},\ \bibinfo {pages} {229} (\bibinfo {year} {2010})},\ \Eprint
  {http://arxiv.org/abs/0910.2191} {arXiv:0910.2191 [nucl-th]} \BibitemShut
  {NoStop}%
%%CITATION = ARXIV:0910.2191;%%
\bibitem [{\citenamefont {Kreuzer}\ and\ \citenamefont
  {Grie{\ss}hammer}(2012)}]{Kreuzer:2012sr}%
  \BibitemOpen
  \bibfield  {author} {\bibinfo {author} {\bibfnamefont {S.}~\bibnamefont
  {Kreuzer}}\ and\ \bibinfo {author} {\bibfnamefont {H.~W.}\ \bibnamefont
  {Grie{\ss}hammer}},\ }\href {\doibase 10.1140/epja/i2012-12093-6} {\bibfield
  {journal} {\bibinfo  {journal} {Eur. Phys. J.}\ }\textbf {\bibinfo {volume}
  {A48}},\ \bibinfo {pages} {93} (\bibinfo {year} {2012})},\ \Eprint
  {http://arxiv.org/abs/1205.0277} {arXiv:1205.0277 [nucl-th]} \BibitemShut
  {NoStop}%
%%CITATION = ARXIV:1205.0277;%%
\bibitem [{\citenamefont {Polejaeva}\ and\ \citenamefont
  {Rusetsky}(2012)}]{Polejaeva:2012ut}%
  \BibitemOpen
  \bibfield  {author} {\bibinfo {author} {\bibfnamefont {K.}~\bibnamefont
  {Polejaeva}}\ and\ \bibinfo {author} {\bibfnamefont {A.}~\bibnamefont
  {Rusetsky}},\ }\href {\doibase 10.1140/epja/i2012-12067-8} {\bibfield
  {journal} {\bibinfo  {journal} {Eur. Phys. J.}\ }\textbf {\bibinfo {volume}
  {A48}},\ \bibinfo {pages} {67} (\bibinfo {year} {2012})},\ \Eprint
  {http://arxiv.org/abs/1203.1241} {arXiv:1203.1241 [hep-lat]} \BibitemShut
  {NoStop}%
%%CITATION = ARXIV:1203.1241;%%
\bibitem [{\citenamefont {Brice{\~n}o}\ and\ \citenamefont
  {Davoudi}(2013{\natexlab{b}})}]{Briceno:2012rv}%
  \BibitemOpen
  \bibfield  {author} {\bibinfo {author} {\bibfnamefont {R.~A.}\ \bibnamefont
  {Brice{\~n}o}}\ and\ \bibinfo {author} {\bibfnamefont {Z.}~\bibnamefont
  {Davoudi}},\ }\href {\doibase 10.1103/PhysRevD.87.094507} {\bibfield
  {journal} {\bibinfo  {journal} {Phys. Rev.}\ }\textbf {\bibinfo {volume}
  {D87}},\ \bibinfo {pages} {094507} (\bibinfo {year} {2013}{\natexlab{b}})},\
  \Eprint {http://arxiv.org/abs/1212.3398} {arXiv:1212.3398 [hep-lat]}
  \BibitemShut {NoStop}%
%%CITATION = ARXIV:1212.3398;%%
\bibitem [{\citenamefont {Hansen}\ and\ \citenamefont
  {Sharpe}(2014)}]{Hansen:2014eka}%
  \BibitemOpen
  \bibfield  {author} {\bibinfo {author} {\bibfnamefont {M.~T.}\ \bibnamefont
  {Hansen}}\ and\ \bibinfo {author} {\bibfnamefont {S.~R.}\ \bibnamefont
  {Sharpe}},\ }\href {\doibase 10.1103/PhysRevD.90.116003} {\bibfield
  {journal} {\bibinfo  {journal} {Phys. Rev.}\ }\textbf {\bibinfo {volume}
  {D90}},\ \bibinfo {pages} {116003} (\bibinfo {year} {2014})},\ \Eprint
  {http://arxiv.org/abs/1408.5933} {arXiv:1408.5933 [hep-lat]} \BibitemShut
  {NoStop}%
%%CITATION = ARXIV:1408.5933;%%
\bibitem [{\citenamefont {Hansen}\ and\ \citenamefont
  {Sharpe}(2015)}]{Hansen:2015zga}%
  \BibitemOpen
  \bibfield  {author} {\bibinfo {author} {\bibfnamefont {M.~T.}\ \bibnamefont
  {Hansen}}\ and\ \bibinfo {author} {\bibfnamefont {S.~R.}\ \bibnamefont
  {Sharpe}},\ }\href {\doibase 10.1103/PhysRevD.92.114509} {\bibfield
  {journal} {\bibinfo  {journal} {Phys. Rev.}\ }\textbf {\bibinfo {volume}
  {D92}},\ \bibinfo {pages} {114509} (\bibinfo {year} {2015})},\ \Eprint
  {http://arxiv.org/abs/1504.04248} {arXiv:1504.04248 [hep-lat]} \BibitemShut
  {NoStop}%
%%CITATION = ARXIV:1504.04248;%%
\bibitem [{\citenamefont {Hansen}\ and\ \citenamefont
  {Sharpe}(2016)}]{Hansen:2016fzj}%
  \BibitemOpen
  \bibfield  {author} {\bibinfo {author} {\bibfnamefont {M.~T.}\ \bibnamefont
  {Hansen}}\ and\ \bibinfo {author} {\bibfnamefont {S.~R.}\ \bibnamefont
  {Sharpe}},\ }\href {\doibase 10.1103/PhysRevD.96.039901,
  10.1103/PhysRevD.93.096006} {\bibfield  {journal} {\bibinfo  {journal} {Phys.
  Rev.}\ }\textbf {\bibinfo {volume} {D93}},\ \bibinfo {pages} {096006}
  (\bibinfo {year} {2016})},\ \bibinfo {note} {[Erratum: Phys.
  Rev.D96,no.3,039901(2017)]},\ \Eprint {http://arxiv.org/abs/1602.00324}
  {arXiv:1602.00324 [hep-lat]} \BibitemShut {NoStop}%
%%CITATION = ARXIV:1602.00324;%%
\bibitem [{\citenamefont {Brice{\~n}o}\ \emph {et~al.}(2017)\citenamefont
  {Brice{\~n}o}, \citenamefont {Hansen},\ and\ \citenamefont
  {Sharpe}}]{Briceno:2017tce}%
  \BibitemOpen
  \bibfield  {author} {\bibinfo {author} {\bibfnamefont {R.~A.}\ \bibnamefont
  {Brice{\~n}o}}, \bibinfo {author} {\bibfnamefont {M.~T.}\ \bibnamefont
  {Hansen}}, \ and\ \bibinfo {author} {\bibfnamefont {S.~R.}\ \bibnamefont
  {Sharpe}},\ }\href {\doibase 10.1103/PhysRevD.95.074510} {\bibfield
  {journal} {\bibinfo  {journal} {Phys. Rev.}\ }\textbf {\bibinfo {volume}
  {D95}},\ \bibinfo {pages} {074510} (\bibinfo {year} {2017})},\ \Eprint
  {http://arxiv.org/abs/1701.07465} {arXiv:1701.07465 [hep-lat]} \BibitemShut
  {NoStop}%
%%CITATION = ARXIV:1701.07465;%%
\bibitem [{\citenamefont {Hammer}\ \emph
  {et~al.}(2017{\natexlab{b}})\citenamefont {Hammer}, \citenamefont {Pang},\
  and\ \citenamefont {Rusetsky}}]{Hammer:2017uqm}%
  \BibitemOpen
  \bibfield  {author} {\bibinfo {author} {\bibfnamefont {H.-W.}\ \bibnamefont
  {Hammer}}, \bibinfo {author} {\bibfnamefont {J.-Y.}\ \bibnamefont {Pang}}, \
  and\ \bibinfo {author} {\bibfnamefont {A.}~\bibnamefont {Rusetsky}},\ }\href
  {\doibase 10.1007/JHEP09(2017)109} {\bibfield  {journal} {\bibinfo  {journal}
  {JHEP}\ }\textbf {\bibinfo {volume} {09}},\ \bibinfo {pages} {109} (\bibinfo
  {year} {2017}{\natexlab{b}})},\ \Eprint {http://arxiv.org/abs/1706.07700}
  {arXiv:1706.07700 [hep-lat]} \BibitemShut {NoStop}%
%%CITATION = ARXIV:1706.07700;%%
\bibitem [{\citenamefont {Hammer}\ \emph
  {et~al.}(2017{\natexlab{c}})\citenamefont {Hammer}, \citenamefont {Pang},\
  and\ \citenamefont {Rusetsky}}]{Hammer:2017kms}%
  \BibitemOpen
  \bibfield  {author} {\bibinfo {author} {\bibfnamefont {H.~W.}\ \bibnamefont
  {Hammer}}, \bibinfo {author} {\bibfnamefont {J.~Y.}\ \bibnamefont {Pang}}, \
  and\ \bibinfo {author} {\bibfnamefont {A.}~\bibnamefont {Rusetsky}},\ }\href
  {\doibase 10.1007/JHEP10(2017)115} {\bibfield  {journal} {\bibinfo  {journal}
  {JHEP}\ }\textbf {\bibinfo {volume} {10}},\ \bibinfo {pages} {115} (\bibinfo
  {year} {2017}{\natexlab{c}})},\ \Eprint {http://arxiv.org/abs/1707.02176}
  {arXiv:1707.02176 [hep-lat]} \BibitemShut {NoStop}%
%%CITATION = ARXIV:1707.02176;%%
\bibitem [{\citenamefont {Mei{\ss}ner}\ \emph {et~al.}(2015)\citenamefont
  {Mei{\ss}ner}, \citenamefont {R{\'\i}os},\ and\ \citenamefont
  {Rusetsky}}]{Meissner:2014dea}%
  \BibitemOpen
  \bibfield  {author} {\bibinfo {author} {\bibfnamefont {U.-G.}\ \bibnamefont
  {Mei{\ss}ner}}, \bibinfo {author} {\bibfnamefont {G.}~\bibnamefont
  {R{\'\i}os}}, \ and\ \bibinfo {author} {\bibfnamefont {A.}~\bibnamefont
  {Rusetsky}},\ }\href {\doibase 10.1103/PhysRevLett.117.069902,
  10.1103/PhysRevLett.114.091602} {\bibfield  {journal} {\bibinfo  {journal}
  {Phys. Rev. Lett.}\ }\textbf {\bibinfo {volume} {114}},\ \bibinfo {pages}
  {091602} (\bibinfo {year} {2015})},\ \bibinfo {note} {[Erratum: Phys. Rev.
  Lett.117,no.6,069902(2016)]},\ \Eprint {http://arxiv.org/abs/1412.4969}
  {arXiv:1412.4969 [hep-lat]} \BibitemShut {NoStop}%
%%CITATION = ARXIV:1412.4969;%%
\bibitem [{\citenamefont {Mai}\ and\ \citenamefont
  {D{\"o}ring}(2017)}]{Mai:2017bge}%
  \BibitemOpen
  \bibfield  {author} {\bibinfo {author} {\bibfnamefont {M.}~\bibnamefont
  {Mai}}\ and\ \bibinfo {author} {\bibfnamefont {M.}~\bibnamefont
  {D{\"o}ring}},\ }\href {\doibase 10.1140/epja/i2017-12440-1} {\bibfield
  {journal} {\bibinfo  {journal} {Eur. Phys. J.}\ }\textbf {\bibinfo {volume}
  {A53}},\ \bibinfo {pages} {240} (\bibinfo {year} {2017})},\ \Eprint
  {http://arxiv.org/abs/1709.08222} {arXiv:1709.08222 [hep-lat]} \BibitemShut
  {NoStop}%
%%CITATION = ARXIV:1709.08222;%%
\bibitem [{\citenamefont {Mai}\ and\ \citenamefont
  {Döring}(2019)}]{Mai:2018djl}%
  \BibitemOpen
  \bibfield  {author} {\bibinfo {author} {\bibfnamefont {M.}~\bibnamefont
  {Mai}}\ and\ \bibinfo {author} {\bibfnamefont {M.}~\bibnamefont {Döring}},\
  }\href {\doibase 10.1103/PhysRevLett.122.062503} {\bibfield  {journal}
  {\bibinfo  {journal} {Phys. Rev. Lett.}\ }\textbf {\bibinfo {volume} {122}},\
  \bibinfo {pages} {062503} (\bibinfo {year} {2019})},\ \Eprint
  {http://arxiv.org/abs/1807.04746} {arXiv:1807.04746 [hep-lat]} \BibitemShut
  {NoStop}%
%%CITATION = ARXIV:1807.04746;%%
\bibitem [{\citenamefont {D{\"o}ring}\ \emph {et~al.}(2018)\citenamefont
  {D{\"o}ring}, \citenamefont {Hammer}, \citenamefont {Mai}, \citenamefont
  {Pang}, \citenamefont {Rusetsky},\ and\ \citenamefont {Wu}}]{Doring:2018xxx}%
  \BibitemOpen
  \bibfield  {author} {\bibinfo {author} {\bibfnamefont {M.}~\bibnamefont
  {D{\"o}ring}}, \bibinfo {author} {\bibfnamefont {H.~W.}\ \bibnamefont
  {Hammer}}, \bibinfo {author} {\bibfnamefont {M.}~\bibnamefont {Mai}},
  \bibinfo {author} {\bibfnamefont {J.~Y.}\ \bibnamefont {Pang}}, \bibinfo
  {author} {\bibfnamefont {{\S}.~A.}\ \bibnamefont {Rusetsky}}, \ and\ \bibinfo
  {author} {\bibfnamefont {J.}~\bibnamefont {Wu}},\ }\href {\doibase
  10.1103/PhysRevD.97.114508} {\bibfield  {journal} {\bibinfo  {journal} {Phys.
  Rev.}\ }\textbf {\bibinfo {volume} {D97}},\ \bibinfo {pages} {114508}
  (\bibinfo {year} {2018})},\ \Eprint {http://arxiv.org/abs/1802.03362}
  {arXiv:1802.03362 [hep-lat]} \BibitemShut {NoStop}%
%%CITATION = ARXIV:1802.03362;%%
\bibitem [{\citenamefont {Romero-López}\ \emph {et~al.}(2018)\citenamefont
  {Romero-López}, \citenamefont {Rusetsky},\ and\ \citenamefont
  {Urbach}}]{Romero-Lopez:2018rcb}%
  \BibitemOpen
  \bibfield  {author} {\bibinfo {author} {\bibfnamefont {F.}~\bibnamefont
  {Romero-López}}, \bibinfo {author} {\bibfnamefont {A.}~\bibnamefont
  {Rusetsky}}, \ and\ \bibinfo {author} {\bibfnamefont {C.}~\bibnamefont
  {Urbach}},\ }\href {\doibase 10.1140/epjc/s10052-018-6325-8} {\bibfield
  {journal} {\bibinfo  {journal} {Eur. Phys. J.}\ }\textbf {\bibinfo {volume}
  {C78}},\ \bibinfo {pages} {846} (\bibinfo {year} {2018})},\ \Eprint
  {http://arxiv.org/abs/1806.02367} {arXiv:1806.02367 [hep-lat]} \BibitemShut
  {NoStop}%
%%CITATION = ARXIV:1806.02367;%%
\bibitem [{\citenamefont {Guo}(2017)}]{Guo:2016fgl}%
  \BibitemOpen
  \bibfield  {author} {\bibinfo {author} {\bibfnamefont {P.}~\bibnamefont
  {Guo}},\ }\href {\doibase 10.1103/PhysRevD.95.054508} {\bibfield  {journal}
  {\bibinfo  {journal} {Phys. Rev.}\ }\textbf {\bibinfo {volume} {D95}},\
  \bibinfo {pages} {054508} (\bibinfo {year} {2017})},\ \Eprint
  {http://arxiv.org/abs/1607.03184} {arXiv:1607.03184 [hep-lat]} \BibitemShut
  {NoStop}%
%%CITATION = ARXIV:1607.03184;%%
\bibitem [{\citenamefont {Guo}\ and\ \citenamefont
  {Gasparian}(2017)}]{Guo:2017ism}%
  \BibitemOpen
  \bibfield  {author} {\bibinfo {author} {\bibfnamefont {P.}~\bibnamefont
  {Guo}}\ and\ \bibinfo {author} {\bibfnamefont {V.}~\bibnamefont
  {Gasparian}},\ }\href {\doibase 10.1016/j.physletb.2017.10.009} {\bibfield
  {journal} {\bibinfo  {journal} {Phys. Lett.}\ }\textbf {\bibinfo {volume}
  {B774}},\ \bibinfo {pages} {441} (\bibinfo {year} {2017})},\ \Eprint
  {http://arxiv.org/abs/1701.00438} {arXiv:1701.00438 [hep-lat]} \BibitemShut
  {NoStop}%
%%CITATION = ARXIV:1701.00438;%%
\bibitem [{\citenamefont {Guo}\ and\ \citenamefont
  {Gasparian}(2018)}]{Guo:2017crd}%
  \BibitemOpen
  \bibfield  {author} {\bibinfo {author} {\bibfnamefont {P.}~\bibnamefont
  {Guo}}\ and\ \bibinfo {author} {\bibfnamefont {V.}~\bibnamefont
  {Gasparian}},\ }\href {\doibase 10.1103/PhysRevD.97.014504} {\bibfield
  {journal} {\bibinfo  {journal} {Phys. Rev.}\ }\textbf {\bibinfo {volume}
  {D97}},\ \bibinfo {pages} {014504} (\bibinfo {year} {2018})},\ \Eprint
  {http://arxiv.org/abs/1709.08255} {arXiv:1709.08255 [hep-lat]} \BibitemShut
  {NoStop}%
%%CITATION = ARXIV:1709.08255;%%
\bibitem [{\citenamefont {Guo}\ and\ \citenamefont
  {Morris}(2019)}]{Guo:2018xbv}%
  \BibitemOpen
  \bibfield  {author} {\bibinfo {author} {\bibfnamefont {P.}~\bibnamefont
  {Guo}}\ and\ \bibinfo {author} {\bibfnamefont {T.}~\bibnamefont {Morris}},\
  }\href {\doibase 10.1103/PhysRevD.99.014501} {\bibfield  {journal} {\bibinfo
  {journal} {Phys. Rev.}\ }\textbf {\bibinfo {volume} {D99}},\ \bibinfo {pages}
  {014501} (\bibinfo {year} {2019})},\ \Eprint
  {http://arxiv.org/abs/1808.07397} {arXiv:1808.07397 [hep-lat]} \BibitemShut
  {NoStop}%
%%CITATION = ARXIV:1808.07397;%%
\bibitem [{\citenamefont {Blanton}\ \emph {et~al.}(2019)\citenamefont
  {Blanton}, \citenamefont {Romero-López},\ and\ \citenamefont
  {Sharpe}}]{Blanton:2019igq}%
  \BibitemOpen
  \bibfield  {author} {\bibinfo {author} {\bibfnamefont {T.~D.}\ \bibnamefont
  {Blanton}}, \bibinfo {author} {\bibfnamefont {F.}~\bibnamefont
  {Romero-López}}, \ and\ \bibinfo {author} {\bibfnamefont {S.~R.}\
  \bibnamefont {Sharpe}},\ }\href {\doibase 10.1007/JHEP03(2019)106} {\bibfield
   {journal} {\bibinfo  {journal} {JHEP}\ }\textbf {\bibinfo {volume} {03}},\
  \bibinfo {pages} {106} (\bibinfo {year} {2019})},\ \Eprint
  {http://arxiv.org/abs/1901.07095} {arXiv:1901.07095 [hep-lat]} \BibitemShut
  {NoStop}%
%%CITATION = ARXIV:1901.07095;%%
\bibitem [{\citenamefont {Romero-López}\ \emph {et~al.}(2019)\citenamefont
  {Romero-López}, \citenamefont {Sharpe}, \citenamefont {Blanton},
  \citenamefont {Briceño},\ and\ \citenamefont
  {Hansen}}]{Romero-Lopez:2019qrt}%
  \BibitemOpen
  \bibfield  {author} {\bibinfo {author} {\bibfnamefont {F.}~\bibnamefont
  {Romero-López}}, \bibinfo {author} {\bibfnamefont {S.~R.}\ \bibnamefont
  {Sharpe}}, \bibinfo {author} {\bibfnamefont {T.~D.}\ \bibnamefont {Blanton}},
  \bibinfo {author} {\bibfnamefont {R.~A.}\ \bibnamefont {Briceño}}, \ and\
  \bibinfo {author} {\bibfnamefont {M.~T.}\ \bibnamefont {Hansen}},\ }\href
  {\doibase 10.1007/JHEP10(2019)007} {\bibfield  {journal} {\bibinfo  {journal}
  {JHEP}\ }\textbf {\bibinfo {volume} {10}},\ \bibinfo {pages} {007} (\bibinfo
  {year} {2019})},\ \Eprint {http://arxiv.org/abs/1908.02411} {arXiv:1908.02411
  [hep-lat]} \BibitemShut {NoStop}%
\bibitem [{\citenamefont {Blanton}\ \emph {et~al.}(2020)\citenamefont
  {Blanton}, \citenamefont {Romero-López},\ and\ \citenamefont
  {Sharpe}}]{Blanton:2019vdk}%
  \BibitemOpen
  \bibfield  {author} {\bibinfo {author} {\bibfnamefont {T.~D.}\ \bibnamefont
  {Blanton}}, \bibinfo {author} {\bibfnamefont {F.}~\bibnamefont
  {Romero-López}}, \ and\ \bibinfo {author} {\bibfnamefont {S.~R.}\
  \bibnamefont {Sharpe}},\ }\href {\doibase 10.1103/PhysRevLett.124.032001}
  {\bibfield  {journal} {\bibinfo  {journal} {Phys. Rev. Lett.}\ }\textbf
  {\bibinfo {volume} {124}},\ \bibinfo {pages} {032001} (\bibinfo {year}
  {2020})},\ \Eprint {http://arxiv.org/abs/1909.02973} {arXiv:1909.02973
  [hep-lat]} \BibitemShut {NoStop}%
\bibitem [{\citenamefont {Mai}\ \emph {et~al.}(2020)\citenamefont {Mai},
  \citenamefont {Döring}, \citenamefont {Culver},\ and\ \citenamefont
  {Alexandru}}]{Mai:2019fba}%
  \BibitemOpen
  \bibfield  {author} {\bibinfo {author} {\bibfnamefont {M.}~\bibnamefont
  {Mai}}, \bibinfo {author} {\bibfnamefont {M.}~\bibnamefont {Döring}},
  \bibinfo {author} {\bibfnamefont {C.}~\bibnamefont {Culver}}, \ and\ \bibinfo
  {author} {\bibfnamefont {A.}~\bibnamefont {Alexandru}},\ }\href {\doibase
  10.1103/PhysRevD.101.054510} {\bibfield  {journal} {\bibinfo  {journal}
  {Phys. Rev. D}\ }\textbf {\bibinfo {volume} {101}},\ \bibinfo {pages}
  {054510} (\bibinfo {year} {2020})},\ \Eprint
  {http://arxiv.org/abs/1909.05749} {arXiv:1909.05749 [hep-lat]} \BibitemShut
  {NoStop}%
\bibitem [{\citenamefont {Guo}\ \emph {et~al.}(2018)\citenamefont {Guo},
  \citenamefont {Döring},\ and\ \citenamefont {Szczepaniak}}]{Guo:2018ibd}%
  \BibitemOpen
  \bibfield  {author} {\bibinfo {author} {\bibfnamefont {P.}~\bibnamefont
  {Guo}}, \bibinfo {author} {\bibfnamefont {M.}~\bibnamefont {Döring}}, \ and\
  \bibinfo {author} {\bibfnamefont {A.~P.}\ \bibnamefont {Szczepaniak}},\
  }\href {\doibase 10.1103/PhysRevD.98.094502} {\bibfield  {journal} {\bibinfo
  {journal} {Phys. Rev.}\ }\textbf {\bibinfo {volume} {D98}},\ \bibinfo {pages}
  {094502} (\bibinfo {year} {2018})},\ \Eprint
  {http://arxiv.org/abs/1810.01261} {arXiv:1810.01261 [hep-lat]} \BibitemShut
  {NoStop}%
%%CITATION = ARXIV:1810.01261;%%
\bibitem [{\citenamefont {Guo}(2020{\natexlab{a}})}]{Guo:2019hih}%
  \BibitemOpen
  \bibfield  {author} {\bibinfo {author} {\bibfnamefont {P.}~\bibnamefont
  {Guo}},\ }\href {\doibase 10.1016/j.physletb.2020.135370} {\bibfield
  {journal} {\bibinfo  {journal} {Phys. Lett.}\ }\textbf {\bibinfo {volume}
  {B804}},\ \bibinfo {pages} {135370} (\bibinfo {year} {2020}{\natexlab{a}})},\
  \Eprint {http://arxiv.org/abs/1908.08081} {arXiv:1908.08081 [hep-lat]}
  \BibitemShut {NoStop}%
%%CITATION = ARXIV:1908.08081;%%
\bibitem [{\citenamefont {Guo}\ and\ \citenamefont
  {Döring}(2020)}]{Guo:2019ogp}%
  \BibitemOpen
  \bibfield  {author} {\bibinfo {author} {\bibfnamefont {P.}~\bibnamefont
  {Guo}}\ and\ \bibinfo {author} {\bibfnamefont {M.}~\bibnamefont {Döring}},\
  }\href {\doibase 10.1103/PhysRevD.101.034501} {\bibfield  {journal} {\bibinfo
   {journal} {Phys. Rev.}\ }\textbf {\bibinfo {volume} {D101}},\ \bibinfo
  {pages} {034501} (\bibinfo {year} {2020})},\ \Eprint
  {http://arxiv.org/abs/1910.08624} {arXiv:1910.08624 [hep-lat]} \BibitemShut
  {NoStop}%
%%CITATION = ARXIV:1910.08624;%%
\bibitem [{\citenamefont {Guo}(2020{\natexlab{b}})}]{Guo:2020wbl}%
  \BibitemOpen
  \bibfield  {author} {\bibinfo {author} {\bibfnamefont {P.}~\bibnamefont
  {Guo}},\ }\href {\doibase 10.1103/PhysRevD.101.054512} {\bibfield  {journal}
  {\bibinfo  {journal} {Phys. Rev.}\ }\textbf {\bibinfo {volume} {D101}},\
  \bibinfo {pages} {054512} (\bibinfo {year} {2020}{\natexlab{b}})},\ \Eprint
  {http://arxiv.org/abs/2002.04111} {arXiv:2002.04111 [hep-lat]} \BibitemShut
  {NoStop}%
%%CITATION = ARXIV:2002.04111;%%
\bibitem [{\citenamefont {Guo}\ and\ \citenamefont
  {Long}(2020{\natexlab{a}})}]{Guo:2020kph}%
  \BibitemOpen
  \bibfield  {author} {\bibinfo {author} {\bibfnamefont {P.}~\bibnamefont
  {Guo}}\ and\ \bibinfo {author} {\bibfnamefont {B.}~\bibnamefont {Long}},\
  }\href {\doibase 10.1103/PhysRevD.101.094510} {\bibfield  {journal} {\bibinfo
   {journal} {Phys. Rev. D}\ }\textbf {\bibinfo {volume} {101}},\ \bibinfo
  {pages} {094510} (\bibinfo {year} {2020}{\natexlab{a}})},\ \Eprint
  {http://arxiv.org/abs/2002.09266} {arXiv:2002.09266 [hep-lat]} \BibitemShut
  {NoStop}%
\bibitem [{\citenamefont {Guo}(2020{\natexlab{c}})}]{Guo:2020iep}%
  \BibitemOpen
  \bibfield  {author} {\bibinfo {author} {\bibfnamefont {P.}~\bibnamefont
  {Guo}},\ }\href@noop {} {\  (\bibinfo {year} {2020}{\natexlab{c}})},\ \Eprint
  {http://arxiv.org/abs/2007.04473} {arXiv:2007.04473 [hep-lat]} \BibitemShut
  {NoStop}%
\bibitem [{\citenamefont {Guo}\ and\ \citenamefont
  {Long}(2020{\natexlab{b}})}]{Guo:2020ikh}%
  \BibitemOpen
  \bibfield  {author} {\bibinfo {author} {\bibfnamefont {P.}~\bibnamefont
  {Guo}}\ and\ \bibinfo {author} {\bibfnamefont {B.}~\bibnamefont {Long}},\
  }\href@noop {} {\  (\bibinfo {year} {2020}{\natexlab{b}})},\ \Eprint
  {http://arxiv.org/abs/2007.10895} {arXiv:2007.10895 [hep-lat]} \BibitemShut
  {NoStop}%
\bibitem [{\citenamefont {Hansen}\ \emph {et~al.}(2020)\citenamefont {Hansen},
  \citenamefont {Romero-López},\ and\ \citenamefont
  {Sharpe}}]{Hansen:2020zhy}%
  \BibitemOpen
  \bibfield  {author} {\bibinfo {author} {\bibfnamefont {M.~T.}\ \bibnamefont
  {Hansen}}, \bibinfo {author} {\bibfnamefont {F.}~\bibnamefont
  {Romero-López}}, \ and\ \bibinfo {author} {\bibfnamefont {S.~R.}\
  \bibnamefont {Sharpe}},\ }\href {\doibase 10.1007/JHEP07(2020)047} {\bibfield
   {journal} {\bibinfo  {journal} {JHEP}\ }\textbf {\bibinfo {volume} {07}},\
  \bibinfo {pages} {047} (\bibinfo {year} {2020})},\ \Eprint
  {http://arxiv.org/abs/2003.10974} {arXiv:2003.10974 [hep-lat]} \BibitemShut
  {NoStop}%
\bibitem [{\citenamefont {Fabre de~la Ripelle}(1983)}]{Ripelle:1983hh}%
  \BibitemOpen
  \bibfield  {author} {\bibinfo {author} {\bibfnamefont {M.}~\bibnamefont
  {Fabre de~la Ripelle}},\ }\href {\doibase
  doi.org/10.1016/0003-4916(83)90212-9} {\bibfield  {journal} {\bibinfo
  {journal} {Annals of Physics}\ }\textbf {\bibinfo {volume} {146}},\ \bibinfo
  {pages} {281} (\bibinfo {year} {1983})}\BibitemShut {NoStop}%
\bibitem [{\citenamefont {Fabre de~la Ripelle}(1993)}]{Ripelle:1993gr}%
  \BibitemOpen
  \bibfield  {author} {\bibinfo {author} {\bibfnamefont {M.}~\bibnamefont
  {Fabre de~la Ripelle}},\ }\href {\doibase doi.org/10.1007/BF01344365}
  {\bibfield  {journal} {\bibinfo  {journal} {Few-Body Systems}\ }\textbf
  {\bibinfo {volume} {14}},\ \bibinfo {pages} {1} (\bibinfo {year}
  {1993})}\BibitemShut {NoStop}%
\bibitem [{\citenamefont {Lovelace}(1964)}]{Lovelace:1964mq}%
  \BibitemOpen
  \bibfield  {author} {\bibinfo {author} {\bibfnamefont {C.}~\bibnamefont
  {Lovelace}},\ }\href {\doibase 10.1103/PhysRev.135.B1225} {\bibfield
  {journal} {\bibinfo  {journal} {Phys. Rev.}\ }\textbf {\bibinfo {volume}
  {135}},\ \bibinfo {pages} {B1225} (\bibinfo {year} {1964})}\BibitemShut
  {NoStop}%
\end{thebibliography}%

\end{document}